\documentclass[a4paper]{article}

\usepackage{natbib}
\usepackage{graphicx}
\usepackage[margin=3.5cm]{geometry}
\usepackage{amssymb}
\usepackage{amsmath}
\usepackage{longtable}
\usepackage{pdflscape}
\usepackage{fancyhdr}
\usepackage{url}

\title{The TEAMx Observational Campaign} %% Article title
\author{Manuela Lehner$^1$ \and Claudia Acquistapace$^{2,3}$ \and Marco Arpagaus$^4$ \and Timothy P. Banyard$^5$ \and Francesco Barbano$^6$ \and Kathrin Baumann-Stanzer$^7$ \and Christophe Brun$^8$ \and Warren R. L. Cairns$^9$ \and Charles Chemel$^{10,11}$ \and Helen E. Dacre$^{12}$ \and Paolo di Girolamo$^{13}$ \and Luca di Liberto$^{14}$ \and Giorgio Doglioni$^6$ \and Philipp Gasch$^{15}$ \and Giacomo Gerosa$^{16}$ \and Lorenzo Giovannini$^6$ \and Sonja Gisinger$^{17}$ \and Alexander Gohm$^1$ \and Jan Handwerker$^{15}$ \and Neil P. Hindley$^5$ \and Stefan Kneifel$^{18}$ \and Peter Knippertz$^{15}$ \and Martin Kohler$^{15}$ \and Meinolf Kossmann$^{19}$ \and Stephen Mobbs$^{10,11}$ \and Andrew Orr$^{20}$ \and Andreas Platis$^{21}$ \and Ian Renfrew$^{22}$ \and Didier Ricard$^{23}$ \and Andrew Ross$^{11}$ \and Harald Saathoff$^{24}$ \and Leopold M. Schlagbauer$^1$ \and Stefano Serafin$^{25}$ \and Peter Sheridan$^{26}$ \and Ivana Stiperski$^1$ \and Nadia Vendrame$^{6,a}$ \and Hannes Vogelmann$^{27}$ \and Jutta V\"ullers$^{28}$ \and Helen C. Ward$^1$ \and Clemens Wastl$^{29}$ \and Stephanie Westerhuis$^{1,b}$ \and Andreas Wieser$^{15}$ \and Norman Wildmann$^{17}$ \and G\"unther Z\"angl$^{30}$ \and Dino Zardi$^6$ \and TOC team$^{c}$ \and Mathias W. Rotach$^1$}
\date{}

\begin{document}
\maketitle
% Author affiliation
\begin{itemize} \itemsep-1mm \itshape
    \item[$^1$] Department of Atmospheric and Cryospheric Sciences, University of Innsbruck, Innrain 52f, 6020 Innsbruck, Austria
    \item[$^2$] Institute of Geophysics and Meteorology, University of Cologne, Pohligstra{\ss}e 3, 50969 K\"oln, Germany
    \item[$^3$] Department of Geosciences, University of Padova, Via Giovanni Gradenigo 6, 35131 Padova, Italy
    \item[$^4$] Federal Office of Meteorology and Climatology MeteoSwiss, Operation Center 1, CH-8058 Zurich-Airport, Switzerland
    \item[$^5$] Centre for Climate Adaptation and Environment Research, University of Bath, Claverton Down, Bath, BA2 7AY, United Kingdom
    \item[$^6$] Department of Civil, Environmental and Mechanical Engineering, University of Trento, Via Mesiano 77, 38123 Trento, Italy
    \item[$^7$] Department Environmental Meteorology, GeoSphere Austria, Hohe Warte 38, 1190 Wien, Austria
    \item[$^8$] Laboratoire des \'Ecoulements G\'eophysiques et Industriels, Universit\'e Grenoble Alpes, 1209-1211 rue de la piscine, 38610 Gi\`eres, France
    \item[$^9$] Institute of Polar Sciences, National Research Council (CNR-ISP), Via Torino 155, 30172 Venezia Mestre, Italy
    \item[$^{10}$] National Centre for Atmospheric Science, Fairbairn House, 71-75 Clarendon Road, Leeds, LS2 9PH, United Kingdom
    \item[$^{11}$] School of Earth, Environment and Sustainability, University of Leeds, Woodhouse Lane, Leeds, LS2 9JT, United Kingdom
    \item[$^{12}$] Department of Meteorology, University of Reading, Brian Hoskins Building, Whiteknights Road, Earley Gate, Reading, RG6 6ET, United Kingdom
    \item[$^{13}$] Department of Health Sciences, University of Basilicata, Via dell'Ateneo Lucano 10, 85100 Potenza, Italy
    \item[$^{14}$] Institute of Atmospheric Sciences and Climate, National Research Council (CNR-ISAC), Via Piero Gobetti 101, 40129 Bologna, Italy
    \item[$^{15}$] Institute of Meteorology and Climate Research Troposphere Research, Karlsruhe Institute of Technology, Kaiserstra{\ss}e 12, 76131 Karlsruhe, Germany
    \item[$^{16}$] Department of Mathematics and Physics, Catholic University of the Sacred Heart, Via della Garzetta 48, 25133 Brescia, Italy
    \item[$^{17}$] Institut f\"ur Physik der Atmosph\"are, Deutsches Zentrum f\"ur Luft- und Raumfahrt e.V., M\"unchener Stra{\ss}e 20, Oberpfaffenhofen, Germany
    \item[$^{18}$] Meteorological Institute, Ludwig-Maximilians-University Munich, Theresienstra{\ss}e 37, 80333 M\"unchen, Germany
    \item[$^{19}$] Climate and Environment, Deutscher Wetterdienst (DWD), Frankfurter Stra{\ss}e 135, 63067 Offenbach am Main, Germany
    \item[$^{20}$] British Antarctic Survey, High Cross, Madingley Road, Cambridge, CB3 0ET, United Kingdom
    \item[$^{21}$] Department of Geosciences, Eberhard Karls University of Tuebingen, Geschwister-Scholl-Platz, 72074 T\"ubingen, Germany
    \item[$^{22}$] School of Environmental Sciences, University of East Anglia, Earlham Road, Norwich, NR4 7TJ, United Kingdom
    \item[$^{23}$] University of Toulouse CNRM, M\'et\'eo-France CNRS, 42 Avenue Gaspard Coriolis, 31057 Toulouse, France
    \item[$^{24}$] Institute of Meteorology and Climate Research Atmospheric Aerosol Research, Karlsruhe Institute of Technology, Kaiserstra{\ss}e 12, 76131 Karlsruhe, Germany
    \item[$^{25}$] Department of Meteorology and Geophysics, University of Vienna, Josef-Holaubek-Platz 2, 1090 Wien, Austria
    \item[$^{26}$] Met Office, FitzRoy Road, Exeter, EX1 3PB, United Kingdom
    \item[$^{27}$] Institute of Meteorology and Climate Research Atmospheric Environmental Research, Karlsruhe Institute of Technology, Kreuzeckbahnstra{\ss}e 19, 82467 Garmisch-Partenkirchen, Germany
    \item[$^{28}$] Institute of Meteorology and Climate Research Trace Gases and Remote Sensing, Karlsruhe Institute of Technology, Kaiserstra{\ss}e 12, 76131 Karlsruhe, Germany
    \item[$^{29}$] Regionalstelle Tirol und Vorarlberg, GeoSphere Austria, F\"urstenweg 180, 6020 Innsbruck, Austria
    \item[$^{30}$] Research and Development, Deutscher Wetterdienst (DWD), Frankfurter Stra{\ss}e 135, 63067 Offenbach am Main, Germany
    \item[$^a$] Present affiliation: Research and Innovation Centre, Fondazione Edmund Mach, Via Mach 1, 38098 San Michele All'Adige, Italy
    \item[$^b$] Present affiliation: SRF Schweizer Radio und Fernsehen, Fernsehstrasse 1-4, CH-8052 Z\"urich, Switzerland
    \item[$^c$] Members are listed in \ref{app:coauthors}
\end{itemize}

Corresponding author: Manuela Lehner (manuela.lehner@uibk.ac.at)

%% Abstract
\begin{abstract}
As part of the international research programme TEAMx (multi-scale transport and exchange processes in the atmosphere over mountains---programme and experiment) a one-year long measurement campaign, the TEAMx Observational Campaign (TOC), was conducted between 2024 and 2025 in a north-south transect through the Alps. Building on the dense operational measurement network in the Alps, the TOC was designed to collect long-term atmospheric observations over the highly complex Alpine terrain. During two six-week long Extended Observational Periods, more than 40 research institutions came together to instrument about 30 sites in the four target areas of the TEAMx domain and study different transport processes, from gravity waves to orographic convection, thermally driven flows, and turbulent exchange. In addition to a suite of ground-based in-situ and remote-sensing instruments, observational activities included airborne measurements with up to three research aircraft and multiple UAS. This paper gives an overview of the science goals and the TOC design, together with preliminary analyses that highlight the potential of the collected dataset.
\end{abstract}

%% ----------- MAIN TEXT -----------
%%
%% ---------- INTRODUCTION ----------
\section{Introduction}
\label{sec:introduction}

The total exchange of energy, mass, and momentum  within the atmospheric boundary layer over complex terrain and between the boundary layer and the free atmosphere results from multiple interacting processes  \citep{Lehner2018,Serafin2018}, thus impacting the water and carbon cycles as well as the energy and momentum budgets. While the exchange in the boundary layer over horizontally homogeneous and flat (HHF) surfaces occurs largely in the vertical direction due to turbulent mixing, a number of mountain-induced and mountain-modified flow processes contribute to the exchange in the mountain boundary layer (MoBL), which occur on a wide range of spatio-temporal scales. The international TEAMx (multi-scale Transport and Exchange processes in the Atmosphere over Mountains---programme and eXperiment) research programme has been initiated with the aim of increasing the understanding of the atmospheric processes contributing to the total transport and exchange over complex mountainous terrain, to better represent these processes in numerical models, and to improve associated weather and climate applications \citep{Serafin2020,Rotach2022}. 

For example, thermally driven slope and valley winds have been shown to effectively transport mass and energy \citep[e.g.,][]{Serafin2010,Wagner2015,Leukauf2016}. Slope winds and valley winds in small tributary valleys occur, however, on spatial scales of O(10--100~m) and are thus not resolved in current operational numerical weather prediction (NWP) models. Global NWP models have a horizontal grid spacing of O(10~km) or more, while regional forecasting models use horizontal grid spacings of down to O(1~km) and are starting to go to even higher resolutions. Similarly, regional ensemble climate models can typically afford the resolution of global NWP models \citep[e.g.,][]{Jacob2014} and most advanced ensembles use km-scale grid spacing \citep[e.g.,][]{Ban2021,Pichelli2021}. Processes on smaller, sub-grid scales contributing to the transport and exchange of energy, mass and momentum are parameterised \citep{Chow2019}. This, however, does not include the transport of mass and energy by unresolved thermally driven slope and valley winds.

Lifting processes, either by forced lifting of flow over topography or thermally induced circulations, also contribute to the transport of moisture and thus to moist convective initiation in mountainous terrain \citep{Kirshbaum2018}. An accurate representation of convective initiation in convection parameterisation schemes requires, however, a thorough understanding of the involved processes that cover a range of different scales. These include, for example, the quantification of the transport in small-scale topographic flows and their connection to local terrain features, as well as the role of land-surface and soil moisture characteristics.

The role of gravity-wave drag has long been known and parameterised in NWP models \citep{Palmer1986}. Increasing grid resolutions and the corresponding shift in the partitioning between parameterised and resolved drag, however, threatens to impede model development \citep{Vosper2016, vanNiekerk2023}. This includes not only the gravity-wave drag parameterisation, but also parameterisation schemes for the turbulent orographic form drag and orographic flow blocking.

Together with the large heterogeneity of the surface (soil, land-use, orography), the interactions of these multiple processes result in large spatial heterogeneity and a complex three-dimensional structure of the MoBL \citep{Lehner2018,Lehner2021,Stiperski2025}. Spatially resolved observations of the MoBL structure and, in particular, turbulence characteristics, are, however, rare. The large spatial heterogeneity also means that horizontal turbulent fluxes and advection cannot be neglected, and have been shown, for example, to contribute to the non-closure of the surface-energy balance \citep{Destro2026}. In addition, the boundary-layer approximation, on which all boundary-layer parameterisations are based and for which a theoretical understanding exists \citep{Rotach2025}, can no longer be justified. Even though many HHF-based parameterisations have been shown to be inappropriate for complex terrain \citep[e.g.,][]{Munoz-Esparza2016,Goger2018,Kosovic2020}, they continue to be used in numerical models because of a lack of better alternatives and missing process understanding. From a modelling perspective, exchange parameterisations thus have to be extended to inhomogeneous and sloping terrain for application over mountainous regions.

TEAMx is a bottom-up financed, international research programme. It is a Cross-Cutting Project within WCRP’s (World Climate Research Program) GEWEX (Global Energy and Water Exchanges) Hydroclimatology Panel \citep{Ward2021,Lehner2026} and was an Endorsed Project of WMO’s WWRP (World Weather Research Program) between 2018 and 2024. With more than 40 partner institutions from over 10 countries, institutional and third-party funding of about 20 million euros has been raised for research activities, in addition to institutional crowd funding that has been providing support for a part-time project coordination office.

Details of the various atmospheric processes contributing to the transport in the atmosphere over mountainous terrain, as well as their representation in numerical models of the atmosphere and their relevance for weather and climate services are given in a White Paper \citep{Serafin2020}, which also identifies four main objectives for TEAMx: improved process understanding, joint experiment(s), improving weather and climate models, and support to weather and climate service providers. The TEAMx Observational Campaign (TOC) was designed to address the second objective to collect a unique atmospheric dataset over mountainous terrain. This paper provides an overview of the TOC (Section~\ref{sec:toc}), the deployed instrumentation (Section~\ref{sec:instrumentation}), first research highlights (Sections~\ref{sec:techniques} and \ref{sec:highlights}), and finally a summary and outlook (Section~\ref{sec:summary}).

%% ---------- TOC - DOMAIN AND DURATION ----------
\section{The TEAMx Observational Campaign}
\label{sec:toc}

%% ---------- teamx domain ----------
\subsection{The TEAMx observational domain}
\label{subsec:toc_domain}

\begin{figure}
\centering
\includegraphics[width=\textwidth]{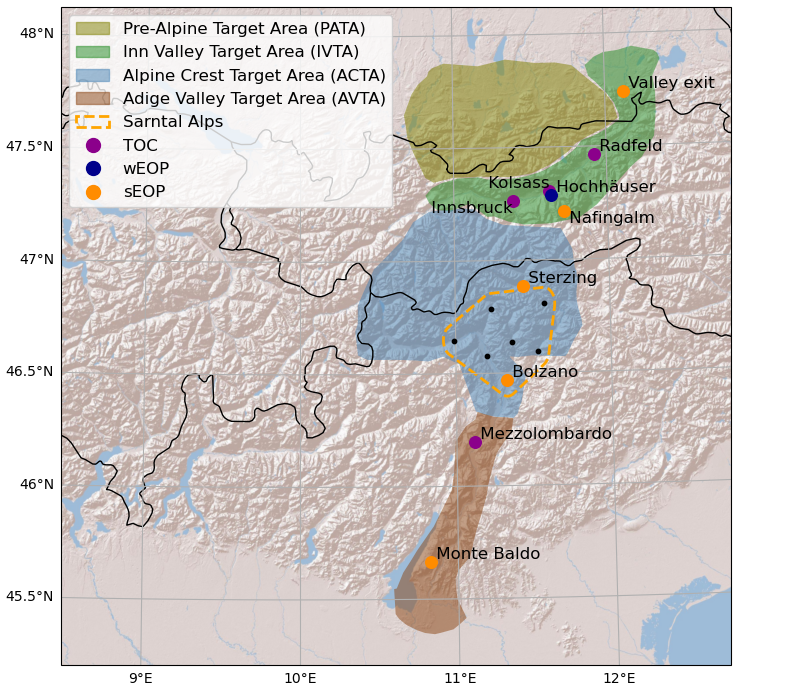}
\caption{The four TEAMx target areas and main measurement sites. Sites instrumented throughout the TOC are shown in purple and sites instrumented during the two Extended Observational Periods in winter (wEOP) and summer (sEOP) in blue and orange, respectively. Instrumentation in the region of the Sarntal Alps (area indicated by the orange dashed line within the ACTA) was distributed across multiple sites (black dots) during the sEOP. Background map: Esri.}
\label{fig:target_areas}
\end{figure}

An approximately north--south transect through the Alps was selected for the TOC (Fig.~\ref{fig:target_areas}) for several scientific and practical reasons \citep{Serafin2020}:
\begin{itemize}
    \item The Alps are an area with extremely complex orography and abundant moisture supply from the Mediterranean Sea and Atlantic Ocean \citep{Sodemann2010}, thus providing ideal conditions to observe moisture transport.
    \item The approximately east--west oriented Inn Valley north of the Alpine crest and the approximately north--south oriented Adige Valley to the south allow the comparison of the energetics of thermally driven flows in two nearby valleys with different orientation under similar synoptic conditions. The two valleys have a comparable geometry (depth, valley floor and crest-to-crest distances) and urban areas of similar dimensions (Innsbruck, Bolzano, and Trento).
    \item The dense routine observational network of automatic weather stations (AWS), eddy-covariance (EC) sites and ceilometers in the observational domain is unparalleled in other mountainous regions of the world.
    \item Extremely diverse surfaces occur over comparably short distances (e.g., urban areas, forests, pastures, lakes, bare rock, glaciated or snow-covered areas). To investigate the boundary-layer structure over glaciated areas, a subgroup of the TEAMx Working Group on Surface-Atmosphere Exchange performed a series of experiments on the Hintereisferner (Hintereisferner Experiment; HEFEX) in summer 2023 \citep{Nicholson2025} and summer 2025 \citep{Schlagbauer2026}, the latter within the framework of the TOC.
    \item The region of the Sarntal Alps (Fig.~\ref{fig:target_areas}) has been identified as  a hotspot for convection initiation \citep{Manzato2022}.
    \item The societal impact of atmospheric research is large because of the high population density and diverse human and economic weather-sensitive activities.
\end{itemize}
% FIGURE: TEAMx domain with target areas and main sites

The observational strategy (Section~\ref{sec:instrumentation}) was designed to observe atmospheric processes across a range of spatial and temporal scales and to capture the multiscale interactions. The observational domain was split into four Target Areas (TAs): the Pre-Alpine TA (PATA), the Inn Valley TA (IVTA), the Alpine Crest TA (ACTA) and the Adige Valley TA (AVTA). With the exception of the PATA, each of the TAs hosted several key measurement sites during two Extended Observation Periods (EOPs, Section~\ref{subsec:toc_eops}) in winter (wEOP) and summer (sEOP) to study one or multiple key processes targeted in TEAMx (Table~\ref{tab:supersites}).
% TABLE: supersites
\begin{table}
\centering
\begin{tabular}{l l l}
    \hline
    Site & Period & Research focus \\
    \hline
    Innsbruck & TOC & MoBL profiling, turbulence \\
    Kolsass & TOC & MoBL profiling, fog, turbulence \\
    Hochh\"auser & wEOP & slope winds, turbulence \\
    Nafingalm & sEOP & MoBL profiling, thermal flows, turbulence \\
    Radfeld & TOC & MoBL profiling, thermal flows, turbulence \\
    Valley exit & sEOP & valley-exit jets \\
    Sarntal Alps & sEOP & MoBL profiling, thermal flows, convection \\
    Bolzano & sEOP & MoBL profiling, thermal flows, convection \\
    Sterzing & wEOP, sEOP & MoBL profiling, convection, gravity waves \\ 
    Mezzolombardo & TOC & turbulence \\
    Monte Baldo & sEOP & slope winds, turbulence \\
    \hline
\end{tabular}
\caption{Key measurement sites and their respective research focus. See Fig.~\ref{fig:target_areas} for site locations.}
\label{tab:supersites}
\end{table}

\begin{figure}
\centering
\includegraphics[width=\textwidth]{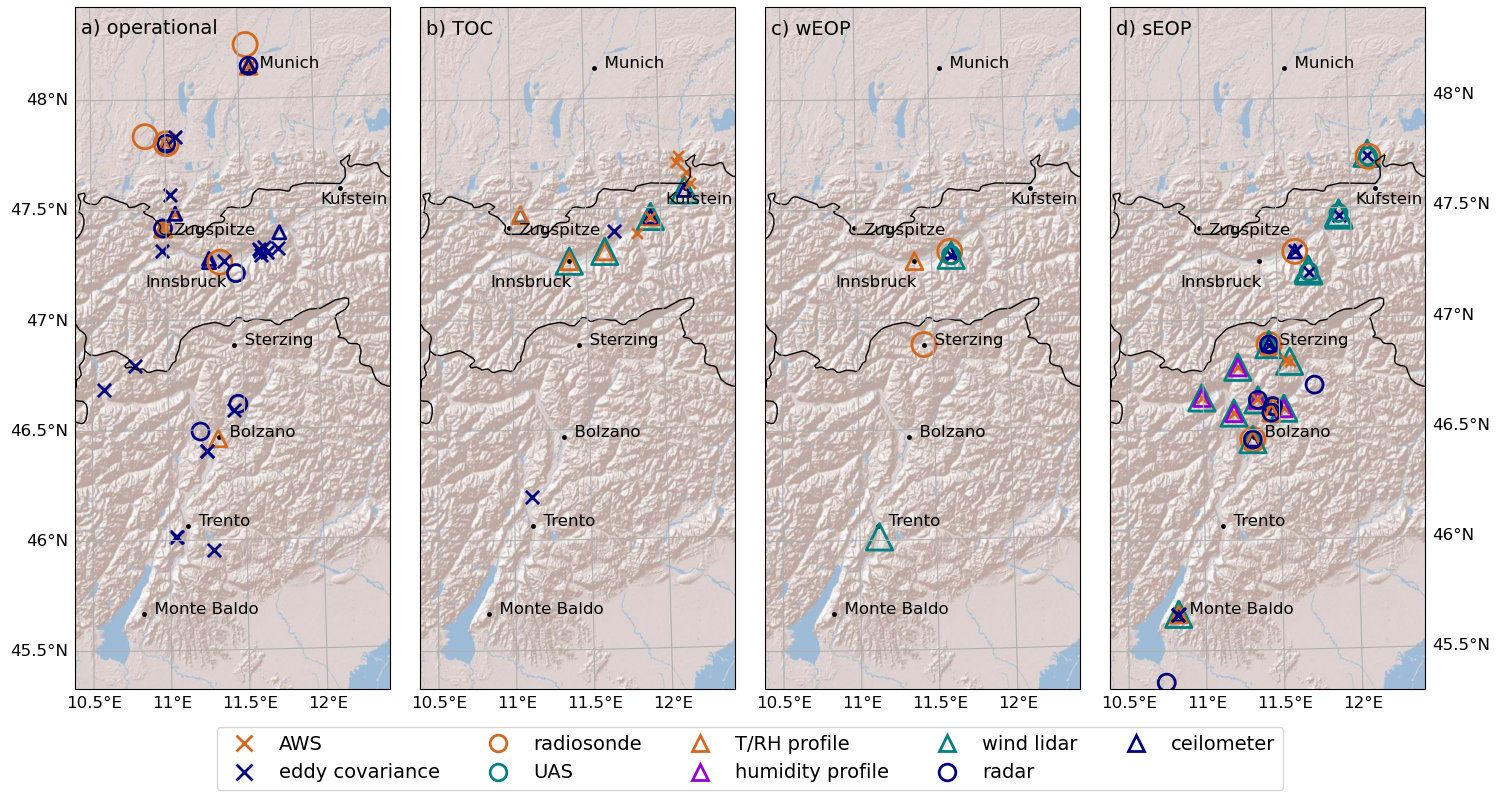}
\caption{(a) Operational instrumentation in the TEAMx observational domain and instrumentation added during the (b) TOC, (c) wEOP, and (d) sEOP. Some of the sites included multiple instruments of the same type which are, however, shown only as one symbol. Eddy-covariance stations typically also included standard AWS instrumentation. (a) does not include operational AWS networks. Background map: Esri}
\label{fig:instrumentation}
\end{figure}

 The TOC lasted for one year from September 2024 to September 2025, during which the existing dense network of operational observations in the TEAMx domain (Fig.~\ref{fig:instrumentation}a) was extended by a number of targeted additional deployments (Fig.~\ref{fig:instrumentation}b). These included, e.g., an additional profiling site with a Doppler wind lidar, a microwave temperature/humidity profiler, and a ceilometer in the Inn Valley, and a complete surface energy balance station on the valley floor of the Adige Valley.
 % FIGURE: instrumentation

%% ---------- eops ----------
\subsection{Extended Observational Periods}
\label{subsec:toc_eops}
During two 6-week long EOPs, one in winter (20~January--28~February 2025, wEOP) and one in summer (16~June--25~July 2025, sEOP), additional instrumentation was deployed (Figs.~\ref{fig:instrumentation}c,d) targeting specific atmospheric processes. Most of the instrumentation ran continuously throughout the EOPs (or beyond). Personnel-intensive measurements such as radiosonde launches, aircraft flights, and UAS operations were conducted during Intensive Observational Periods (IOPs).

During both EOPs daily meetings were held among the participating field scientists to decide about upcoming IOP activities. Weather forecasts were supported by high-resolution NWP runs from multiple European weather services with grid spacings of 500--1.25~km (Table~\ref{tab:nwp}). The high-resolution forecasts provided valuable guidance for the onset of convection and the valley-wind circulation as well as the formation and breakup of low stratus in the valleys. Section~\ref{subsec:techniques_icon} provides further details about the ICON-A05 as an example highlighting the use during the TOC.
% TABLE: NWP
\begin{table}
\centering \footnotesize
\begin{minipage}{\textwidth}
\begin{tabular}{p{1.7cm} p{1.8cm} p{1.2cm} p{0.7cm} p{1.8cm} p{1.5cm} p{1.8cm}}
    \hline
    Weather service & Model & Grid spacing & Lead time & Runs & Periods & Archived \\
    \hline
    DWD & ICON-A05 & 500 m & 48 h & 3-hourly & TOC & 00/12 UTC \\
    Met Office & MetUM & 500 m & 72 h & 12 UTC & Jul--Sep\footnote{some days missing because of technical problems} & yes \\
    Met Office & MetUM & 1 km & 72 h & 12 UTC & Jan--Sep & yes \\
    M\'et\'eo-France & AROME & 500 m & 72 h & 18 UTC & TOC & yes \\
    M\'et\'eo-France & AROME & 1.25 km & 72 h & 00/18 UTC & TOC & yes \\
    Meteo Swiss & ICON-CH1-EPS & 1 km & 45 h & 03 UTC & TOC & subset of control run \\
    GeoSphere Austria & AROME C-LAEF-1k & 1 km & 60 h & 00/12 UTC & TOC & subset of control run \\
    \hline
\end{tabular}
\end{minipage}
\caption{High-resolution NWP runs performed for the TEAMx observational domain during the entire or part of the TOC.}
\label{tab:nwp}
\end{table}

Table~\ref{tab:iops} gives an overview of all IOPs conducted during the wEOP (wIOPs) and the sEOP (sIOPs), with a detailed list of all IOPs in Tables~\ref{tab:iops_weop} and \ref{tab:iops_seop} in the Appendix. IOP types were defined based on the science goals and the available observational resources:
\begin{description}
    \item[SLP] IOPs focused on the local characteristics and spatial variability of the mean-flow and turbulence characteristics in slope flows, specifically in a katabatic flow on an ideal, north-facing slope site in the IVTA during the wEOP and anabatic slope flows on a wide slope of Monte Baldo in the AVTA during the sEOP.
    \item[MoBL] IOPs focused on the three-dimensional structure, transport processes, turbulence characteristics and exchange with the free troposphere. During the wEOP, IOPs were conducted in the IVTA both under stable, thermally driven conditions (MoBL-T) and under the influence of dynamic forcing (MoBL-D). During the sEOP, MoBL-T IOPs were conducted during unstable (convective), thermally driven conditions both in the IVTA and the ACTA.
    \item[VEJ] IOPs were conducted at the exit of the Inn Valley (IVTA) during the sEOP to determine the spatial structure and temporal evolution of nocturnal valley-exit jets and their forcing mechanisms under synoptically undisturbed conditions.
    \item[CON] IOPs focused on convection and convective initiation over the Sarntal Alps in the ACTA during the sEOP, an area which has been identified as a hotspot for orographic convection initiation \citep{Manzato2022}.
    \item[GW] IOPs were conducted during both EOPs to study the gravity-wave development at the Alpine crest and the associated wave propagation.
    \item[SYN:] A synoptically driven event was observed at the end of the sEOP, including a frontal passage over the TEAMx observational domain.
\end{description}
% TABLE: IOP overview
\begin{table}
\centering
\begin{tabular}{l l l l}
    \hline
    IOP type & Abbreviation & wEOP & sEOP \\
    \hline
    Slope winds & SLP & 6 & 2 \\
    Valley-exit jets & VEJ & - & 2 \\
    MoBL-thermal & MoBL-T & 2 & 14 \\
    MoBL-dynamic & MoBL-D & 1 & 1 \\
    Convection & CON & - & 7 \\
    Gravity waves & GW & 9 & 2 \\
    Synoptic & SYN & - & 1 \\
    \hline
    Total & & 17 & 24 \\
    \hline
\end{tabular}
\caption{Number of IOPs for each IOP type. The total number of IOPs deviates from the sum of IOPs per IOP type because some IOPs had multiple foci.}
\label{tab:iops}
\end{table}

%% ---------- synoptic overview -----------
\subsection{Synoptic situation during the wEOP and sEOP}
\label{subsec:synoptic_overview}

Monthly mean temperatures during the wEOP were 2--3~K higher than during the reference period 1990--2020 \citep{GeoSphere2025}, with up to 75\% less precipitation. The wEOP began after a period of relatively persistent high pressure over the Alps, with a sequence of major troughs propagating from north-west over the TOC domain, resulting in varying wind regimes and extended periods with a strong meridional flow component exciting gravity waves. Correspondingly, the first half of the wEOP was dominated by GW IOPs focusing on the large scale and SLP IOPs focusing on the local scale (Table~\ref{tab:iops_weop}). During the second half of the wEOP, the synoptic influence weakened, leading to a number of thermal (or weakly dynamically influenced) MoBL IOPs in the IVTA. Local moisture accumulation in the Inn Valley and low-stratus formation during the nights affected the planning of aircraft operations. Overall, the weather conditions allowed for at least one near-ideal IOP for all science projects during the wEOP . 

The sEOP started during a rather warm (up to 3~K warmer than the reference period) and dry (-30\% total precipitation) June, transitioning to a relatively cool (-1~K) and moist (+20\%) July. The TOC domain remained under the influence of passing ridge and trough axes, with alternating precipitation events and periods of reduced dynamic influence. An exception was the establishment of a relatively persistent, multi-day anti-cyclonic influence at the end of June, leading to a series of MoBL-T, CON, and SLP IOPs. Frontal passages and changing synoptic influence before and after this period led to generally shorter IOPs, which did not always allow observing the full diurnal cycle of the target processes. With the exception of GW IOPs, which were not expected to be abundant during the sEOP, all science projects could realize several near-ideal IOPs.

%% ---------- INSTRUMENTATION ----------
\section{Instrumentation}
\label{sec:instrumentation}

Atmospheric observations in complex terrain suffer from several challenges compared to flat terrain \citep{Emeis2018}, including the need for high spatial measurement density and the need to assess the representativeness of observations. These requirements were addressed for the TOC in that over 40 research groups contributed observational resources (Table~\ref{tab:institutions}), while representativeness can be assessed using the spatially dense observational network and spatially resolving measurements, complemented by high-resolution numerical modelling. A suite of observational devices and techniques was employed to sample the atmosphere from small-scale turbulence to the large-scale flow across the Alps, from the surface layer to the troposphere above the MoBL, and during differing flow conditions. Figure~\ref{fig:instrumentation} shows maps of key measurements within the TEAMx observational domain during the entire TOC, the wEOP, and the sEOP. 

%% ---------- airborne ----------
\subsection{Airborne observations}
\label{subsec:instrumentation_airborne}
\subsubsection{Research aircraft}
\begin{figure}
\centering
\includegraphics[width=0.8\linewidth]{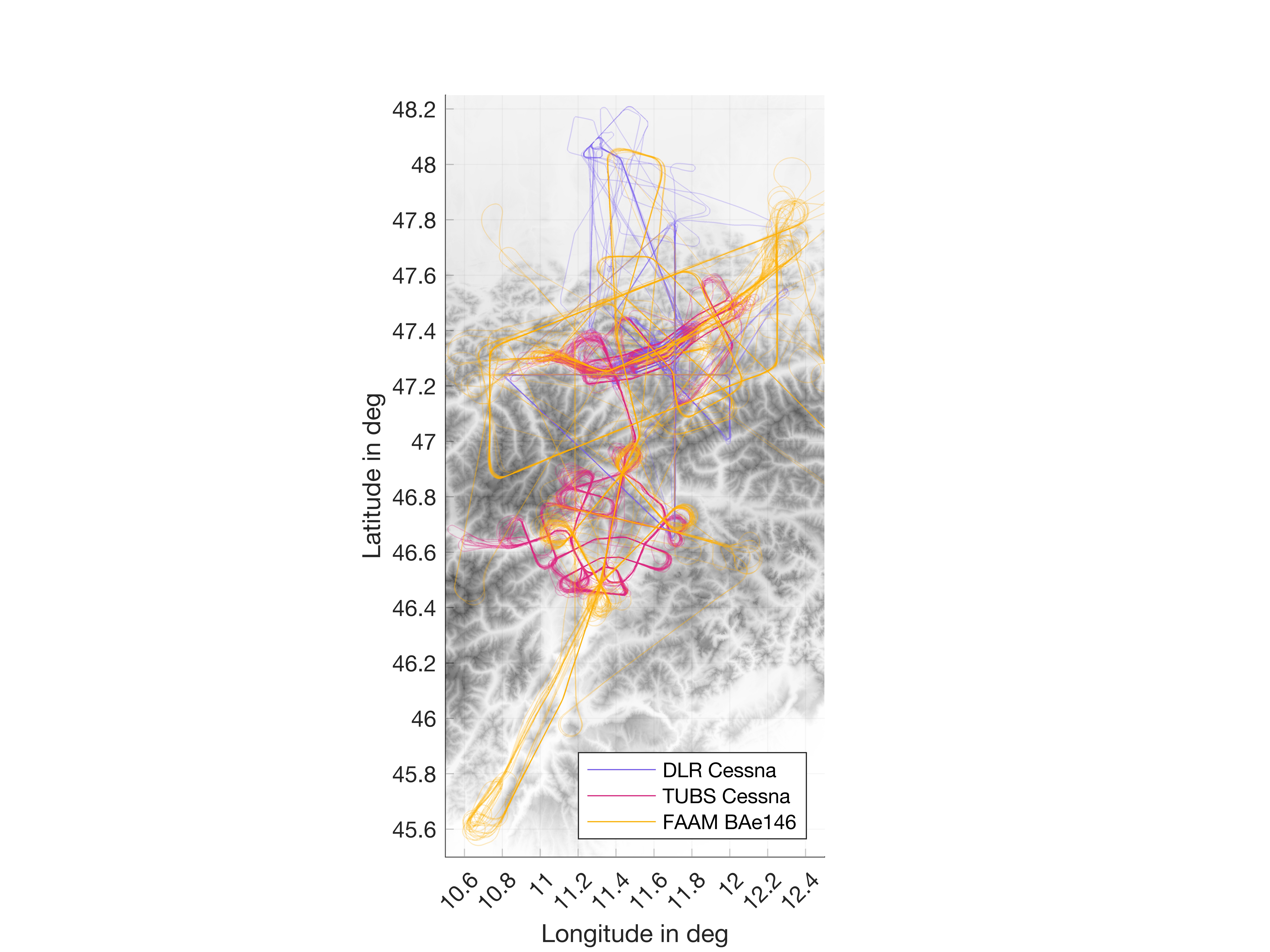}
\caption{Overview of all flight tracks during the sEOP.}
\label{fig:flight_tracks}
\end{figure}

Airborne measurements were conducted with three research aircraft: the FAAM (Facility for Airborne Atmospheric Measurements) BAe146 operated by the National Centre for Atmospheric Science (NCAS), a Cessna F406 operated by the Technische Universit\"at Braunschweig (TUBS), and a Cessna Grand Caravan operated by the German Aerospace Center DLR. During three weeks of the sEOP (7--26~July), all three aircraft participated in the TOC together, with the FAAM and TUBS aircraft stationed at Innsbruck airport and the DLR Cessna operating from Oberpfaffenhofen, Germany. TUBS Cessna flights started already two weeks earlier during the sEOP (from 23 June). Additionally, the DLR Cessna also conducted flights in the IVTA on three days during the wEOP. In total, 265 flight hours were conducted during 80 flights (Table~\ref{tab:flights}). An overview of the flight paths during the sEOP is shown in Fig.~\ref{fig:flight_tracks}, which cover all four target areas with a focus on the IVTA and ACTA.
% TABLE: flight hours
\begin{table}
\centering
\begin{tabular}{p{1.5cm} p{2.9cm} p{1.1cm} p{1.3cm} p{1.1cm} p{3.2cm}}
    \hline
    Aircraft & Core instrumentation & Flight days & Flights & Flight hours & IOPs (TAs) \\
    \hline
    DLR Cessna Caravan & 100-Hz in-situ & 3/10 & 6/18 & 23.9/ 73.6 & MoBL (IVTA), GW (IVTA, ACTA) \\ \hline
    TUBS Cessna F406 & 100-Hz in-situ, AIRflows & 20 & 35 & 97.9 & MoBL (IVTA, ACTA), CON (ACTA) \\ \hline
    FAAM BAe146 & 32-Hz in-situ, dropsondes, aerosol lidar & 20 & 27 & 93.5 & MoBL, GW, CON (all TAs) \\
    \hline
\end{tabular}
\caption{Overview of conducted aircraft missions. Flight statistics include transfer and test flights. Statistics for the DLR Cessna include values for the wEOP (first number) and sEOP (second number).}
\label{tab:flights}
\end{table}
% FIGURE: flight patterns

The DLR Cessna Caravan carried a 100 Hz in-situ probe for 3D wind, humidity and temperature measurements. Operations were conducted within the valley atmosphere below 3000~m~MSL to focus on the spatially variable turbulence structure in the Inn Valley and around 4000--5000~m~MSL to focus on gravity waves and their interactions with low-level processes within the IVTA and ACTA.

The TUBS Cessa F406 also carried a 100 Hz in-situ probe. Additionally, a novel multi-beam Doppler lidar \citep[AIRflows, ][Section~\ref{subsec:techniques_aircraft}]{Gasch2025} developed by KIT provided remotely sensed wind measurements below the aircraft at 100-m along-track resolution. The TUBS Cessna F406 operated typically around 3000--4000~m~MSL to capture mesoscale variations in the mean and turbulent flow. In the IVTA, TUBS and DLR flights were aligned whenever possible to enable validation of the AIRflows mean wind and turbulence measurements and to provide the mesoscale context for turbulence measurements by the DLR Cessna. The TUBS Cessna and AIRflows system were additionally operated in the ACTA, focusing on valley flows and their relation to vertical exchange. Repeated cross-valley transects were flown to validate ground-based valley volume flux estimates from the KITcube lidar network (Section~\ref{subsec:instrumentation_profiling}). In addition, cross-mountain range transects linked valley flows and vertical exchange across the Sarntal Alps mountain range.

The FAAM BAe146 carried in-situ measurements (high-frequency 3D wind, temperature, multiple humidity sensors, short- and long-wave radiation, chemistry (CO, SO$_{\textnormal{2}}$, O$_{\textnormal{3}}$), aerosols (Passive Cavity Aerosol Spectrometer Probe), and cloud microphysics (cloud droplet probe, cloud imaging probes, cloud-aerosol and precipitation spectrometer)), a dropsonde system, as well as an aerosol backscatter lidar (Leosphere ALS450). The BAe146 operated at mid- to upper levels across the full TEAMx observational domain. Flights in the IVTA focused on gravity waves and mesoscale exchange processes using a box-pattern for budget estimations. Flights in the ACTA and across the full Alps focused on mesoscale conditions prior to convective initiation as well as micro-physical processes in convective clouds.

\subsubsection{Unmanned aircraft systems}
In addition to the research aircraft covering entire TAs, as well as reaching altitudes above crest level, airborne measurements were conducted using unmanned aircraft systems (UAS). Multiple UAS were used simultaneously at the Nafingalm and Radfeld measurement sites (IVTA) during the sEOP, with a focus on MoBL IOPs (Section~\ref{subsec:techniques_uas}). The SWUF-3D fleet \citep[Simultaneous Wind measurement with Unmanned Flight Systems in 3D; e.g.,][]{Wildmann2025}, consisting of 30 individual multicopters, was strategically deployed to map spatially distributed, sub-mesoscale flow fields at Nafingalm. At Radfeld, four UAS of type MASC-MC (Multi-Purpose Sensor Carrier---MultiCopter) were operated along a transect from the valley floor to the north mountain sidewall to characterize the boundary layer in the Inn Valley via vertical profiles, capturing the mesoscale effects of the flow within the valley. UAS were also used to sample the vertical structure of slope winds during the wEOP and the nighttime valley-exit jet at the exit of the Inn Valley during the sEOP.

%% ---------- profiling ----------
\subsection{Profiling observations}
\label{subsec:instrumentation_profiling}
\subsubsection{Balloon soundings}
Operational radiosonde sites are sparse within the TOC domain, with launches twice per day at Munich, occasional launches at Hohenpeissenberg and at the German military base Altenstadt (PATA). Within the inner Alpine regions of the IVTA, ACTA, and AVTA, Innsbruck airport is the only operational site, with one nighttime launch per day. During both EOPs, additional sondes were launched regularly at noon at Innsbruck airport, the launch interval at Altenstadt was increased to 6~h, and two launches per day were performed at Sterzing (ACTA). During the sEOP, two soundings per day were also performed from Bolzano (ACTA). In addition to this regular monitoring, frequent soundings (usually at 3-h intervals) were performed during IOPs from Kolsass and Sterzing during both EOPs and from Bolzano during the sEOP. During wEOP GW IOPs, two sondes were often launched simultaneously from Sterzing to allow the determination of wave properties (Section~\ref{subsec:techniques_radiosoundings}).

In-situ profiling was also performed using a tethered-balloon system to sample the vertical structure of slope winds up to about 150~m~AGL, for both katabatic winds over a snow-covered slope in the IVTA during the wEOP and the diurnal cycle of summertime slope flows during the sEOP at Monte Baldo (AVTA). One day of the sEOP also saw the release of a cluster of neutral buoyancy sondes from Sterzing to collect data along the Lagrangian path of the sondes \citep{Abdunabiev2024}.

\subsubsection{Remote-sensing instrumentation}
Remote-sensing instrumentation provided the backbone of the TOC, with several sites, either operational or established for the TOC, yielding data throughout the entire year, including vertical profiles of wind, temperature, and humidity (Fig.~\ref{fig:instrumentation}). Vertical temperature profiles were collected using both Raman lidars and microwave temperature profilers. Two Raman lidars \citep[Purple Pulse Lidar Systems;][]{Lange2019} were operated throughout the TOC at Garmisch-Partenkirchen (PATA) and Innsbruck (IVTA). During the sEOP, Raman lidars was also deployed at Monte Baldo \citep[AVTA;
][]{DiGirolamo2023} and at Bolzano. In addition, seven microwave radiometers \citep[HATPRO, Radiometer Physics;][]{Rose2005} were deployed across multiple sites in the ACTA and IVTA. Some of the radiometers were also performing azimuth scans at fixed elevations to derive the spatial variability of total water vapor and temperature.

In total about 25 Doppler wind lidars were deployed during the TOC, ranging from short-range profiling instruments (METEK Wind Ranger) to capture the wind profiles in the lowest few hundred meters above the ground to fully scanning instruments (Leosphere Windcube 200s and 100s, Halo Photonics Streamline, and Lockheed Martin Coherent Technologies WindTracer instruments) to observe spatially distributed wind fields. The scanning strategies for each of the instruments depended on the measurement site and research focus of the respective group, including VAD (velocity azimuth display) and DBS (Doppler beam swinging) scans for vertical profiles of horizontal wind speed and direction, scans in along- and across-valley direction, and Dual Doppler scans to quantify the two-dimensional wind field within the scanning plane.

During the sEOP, seven KITcube \citep{Kalthoff2013} remote-sensing sites were established throughout the Sarntal Alps (ACTA, Figs.~\ref{fig:target_areas} and \ref{fig:instrumentation}c). Each of the sites was instrumented with a Doppler wind lidar (five Leosphere Windcube WLS200s and two Lockheed Martin Coherent Technologies WindTracer) and five of them with a DIAL (Differential Absorption Lidar, Vaisala DA10) for humidity profiles. The sites were additionally equipped with an AWS, cloud camera, Parsivel disdrometer, and other auxiliary measurements. At one station (Sarnthein) a cloud radar was added. Two additional sites in the ACTA were instrumented with a Parsivel disdrometer and a microwave radiometer to capture vertical profiles of temperature and humidity as well as precipitation at different elevations. With a focus on convection initiation in the ACTA, two mountain-top sites were equipped with X-band radars and three sites with a K-Band radar (micro rain radar). Together with the operational radars in the area, dense radar coverage was thus ensured for the region.

\subsubsection{Data distribution}
Some of the profiles were also distributed in near real-time, in particular, radiosonde launches through the WMO Global Telecommunication System (GTS) and some of the remote-sensing profile observations through the E-PROFILE programme of EUMETNET (https://e-profile.eu). The purpose of real-time data sharing was to allow the assimilation of the data in operational forecast models, such as the ICON-A05 runs produced for the TOC by DWD (Section~\ref{subsec:techniques_icon}). 

%% ---------- surface obs ----------
\subsection{Surface observations}
\label{subsec:instrumentation_surface}
\subsubsection{Eddy-covariance stations}
Turbulent transport is among the key processes contributing to the exchange between the near-surface atmosphere and the free troposphere, however, most current surface-atmosphere exchange parameterizations are only valid in HHF conditions \citep{Rotach2025}. Eddy-covariance (EC) is currently the standard technique to observe turbulent motions in the atmosphere, however, each step in the post-processing of EC measurements requires careful assessment over complex terrain \citep[e.g.,][]{Stiperski2016}. Only observational campaigns like the TOC can provide sufficient spatial resolution in turbulence measurements to validate the applicability of the HHF assumptions for complex terrain.

Several semi-operational EC stations are operated within the TEAMx observational domain providing long-term records of EC measurements at different locations, including the i-Box station network in the Inn Valley \citep{Rotach2017}, the Innsbruck Atmospheric Observatory (IAO) in Innsbruck \citep{Karl2020,Ward2022}, the Forest-Atmosphere Interaction Research (FAIR) station west of Innsbruck \citep{Platter2024}, the TERENO Pre-Alpine Observatory \citep{Kiese2018}, eight ICOS \citep[Integrated Carbon Observation System,][]{Heiskanen2022}, and five FLUXNET \citep{Baldocchi2001} stations. Additional long-term sites were established within the framework of TEAMx which will be maintained beyond the duration of the TOC, including a multi-level site at Mezzolombardo in the Adige Valley and a new site added to the i-Box network in the Inn Valley.

During the wEOP, a network of seven EC stations, each equipped with a nanobarometer, and a distributed temperature sensing array were deployed at the slope site Hochh\"auser (IVTA). One of the stations included two multi-hole pitot tubes on a moving platform that allowed measurements of near-surface turbulence down to about 2~mm above the ground. The objective was to capture the turbulence characteristics of the katabatic flow throughout its depth, and its variations along and across the approximately 200-m long and 50-m wide slope.

During the sEOP, three EC stations were deployed at Monte Baldo (AVTA), one including two multi-hole pitot tubes, to observe the diurnal cycle of summertime slope winds. At Kolsass (IVTA), a network of seven EC stations with nanobarometers was deployed along and across the valley from one of the i-Box stations. The network was supplemented by a scintillometer to allow a comparison of spatially averaged turbulence measurements with distributed point measurements. A scintillometer was also deployed at Mezzolombardo (AVTA).

\subsubsection{Surface station networks}
AWS, disdrometer, and microbarograph networks were also deployed at several locations, for example, an AWS network at Nafingalm (IVTA) during the sEOP to capture the variability in surface conditions; temperature and humidity sensor networks at Nafingalm and Monte Baldo (AVTA); and a network of Parsivel disdrometers located at the KITcube remote-sensing sites in the Sarntal Alps. Downstream of the Inn Valley exit, a network of AWS was set up as part of the TEAMx pre-campaign PC22 \citep{Pfister2024} in 2022 to study the spatial extent of nighttime valley-exit jets, which continued to collect data throughout the TOC. Additional EC stations and AWS were deployed at a number of additional locations to supplement other instrumentation. The installations at Monte Baldo (AVTA) and the KITcube main site Bolzano (ACTA) also focused on air chemistry observations during the sEOP, featuring multiple ozone and particulate matter probes along the slope of Monte Baldo as well as measurements of ice nucleating particle (INP) number concentrations, trace gas concentrations, black carbon, and aerosol particle number, size and chemical composition at Bolzano.

%% ---------- TECHNIQUES ----------
\section{First highlights---observational techniques and numerical modeling}
\label{sec:techniques}

%% ---------- aircraft ----------
\subsection{Aircraft measurements to observe the valley atmosphere}
\label{subsec:techniques_aircraft}

Routine observations are usually restricted to near-surface point measurements, for example, from AWS networks. While remote-sensing instrumentation provides additional information about the vertical structure of the atmosphere, measurements are still mostly limited to individual locations and to mean flow properties. On the other hand, observations of the three-dimensional structure of the MoBL, in particular of the turbulence characteristics, are relatively rare. Aircraft measurements can thus yield an important contribution to fill this gap.

\begin{figure}
\centering
\includegraphics[width=\linewidth]{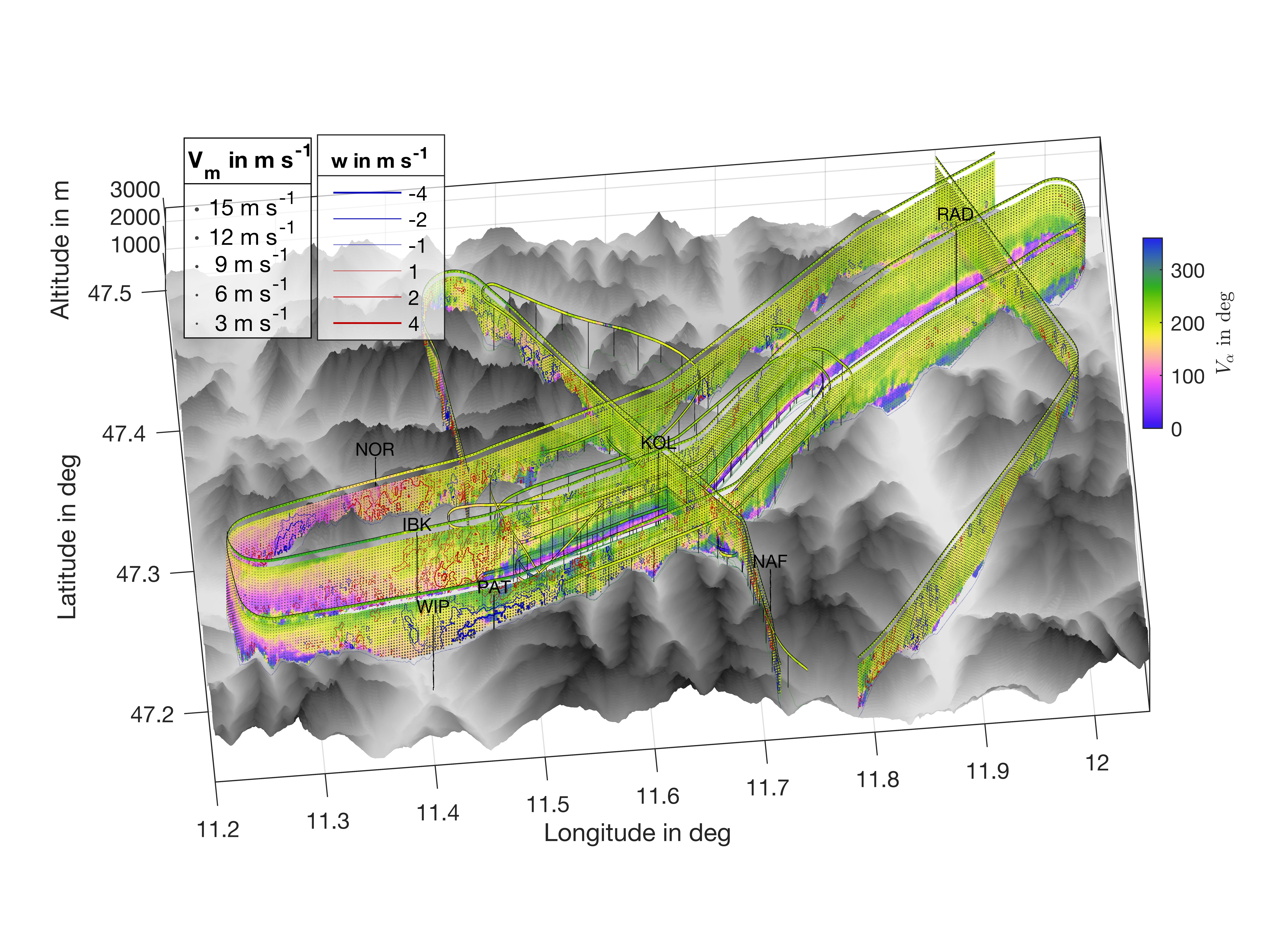}
\caption{DLR (1440--1622~UTC) in-situ and TUBS Cessna (1500--1602~UTC) in-situ and AIRflows measurements conducted within the IVTA during sIOP20 (20~July 2025). Color-coded is the wind direction, the wind speed is indicated through point size, and the AIRflows vertical wind measurements are shown by contours. In-situ data are shown within the narrow band at flight level. Locations marked in the figure (\textit{NOR} Nordkette, \textit{IBK} Innsbruck, \textit{WIP} Wipp Valley, \textit{KOL} Kolsass, \textit{NAF} Nafingalm, \textit{RAD} Radfeld) are discussed in the text.}
\label{fig:highlight_aircraft}
\end{figure}

Figure~\ref{fig:highlight_aircraft} shows DLR and TUBS Cessna measurements conducted within the IVTA during sIOP20. The TUBS Cessna was flying along- and across-valley legs in the Inn Valley between Innsbruck (label \textit{IBK} in Fig.~\ref{fig:highlight_aircraft}) and Radfeld (\textit{RAD}), while the DLR Cessna sampled a shorter valley segment centered around Kolsass (\textit{KOL}) and Nafingalm (\textit{NAF}). AIRflows measurements provide a unique insight into the mesoscale along- and across-valley flow variation. Strong spatial heterogeneity occurs because of the asymmetric exit of a foehn jet from the Wipp Valley (\textit{WIP}). The foehn is strongest on the eastern side of the Wipp Valley, where flow across the Patscherkofel (\textit{PAT}) mountain induces a strong downdraft on the lee (northern) side with values exceeding $-4$~m~s$^{-1}$. Upon entering the Inn Valley, the flow is channeled into the valley and splits into a south-eastern and south-western branch south of the Nordkette (\textit{NOR}), where regions with updraft speeds above 2~m~s$^{-1}$ occur. Along the eastern Inn Valley, south-westerly flow prevails at upper levels, with more westerly directions towards the surface. Below the channeled foehn flow, a shallow layer with weak easterly winds persists.
FIGURE: techniques - aircraft

The agreement between the AIRflows remote sensing and the DLR Cessna in-situ measurements is high, despite temporal sampling differences of more than 30 minutes at individual locations. The DLR Cessna samples stronger turbulent fluctuations due to the higher temporal resolution of the measurements. The agreement highlights the quality of both measurements, as well as the stationarity of the flow during the sampling period. Future combination with additional measurements in the IVTA will enhance the understanding of the complex dynamics during this and other events.

%% ---------- uas ----------
\subsection{Uncrewed Aerial Systems for Atmospheric Research in Complex Terrain}
\label{subsec:techniques_uas}

Atmospheric flow over complex terrain is inherently non-stationary and spatially heterogeneous. Traditional remote sensing instruments (lidar, sodar, radar) rely on assumptions that frequently break down in such environments, while radiosondes drift with the wind, making them unsuitable for capturing the atmospheric state at a specific location. Installing fixed in-situ instrumentation is often logistically cumbersome or impossible. Small UAS offer a crucial, flexible solution to bridge this observational gap. Multicopter UAS, in particular, can be rapidly deployed to probe the ABL at virtually any point in three-dimensional space. Only in recent years have these systems matured sufficiently to deliver high-quality, turbulence-resolving measurements \citep{Wildmann2022,Wetz2023,Schoen2026}.

\begin{figure}
\centering
\includegraphics[width=\linewidth]{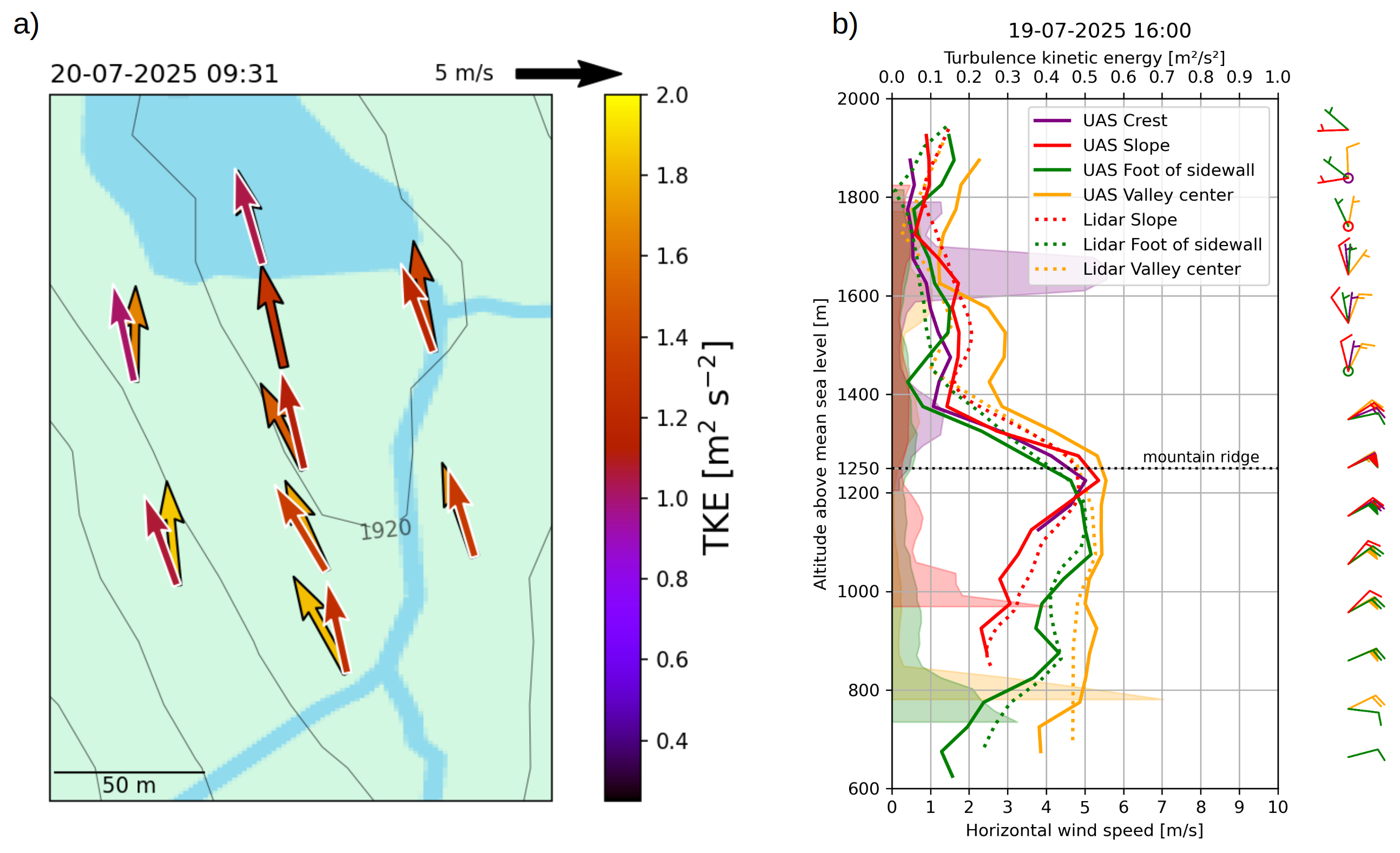}
\caption{(a) SWUF-3D fleet measurements at Nafingalm during sIOP20 (20~July 2025), visualized through wind-speed scaled arrows on the background map. Arrows with black (white) contour lines are measurements at 150~m (50~m) above valley the floor. Arrow color indicates TKE. (b) Observations of the up-valley flow during sIOP19 (19~July 2025, 1600~UTC) at a valley transect close to Radfeld. Solid lines show horizontal wind speed, in the center of the valley (yellow), at the foot of the sidewall (green), at the slope (red) and close to the crest (purple). The wind direction and TKE from the UAS are displayed as wind barbs and shaded areas, respectively, in (b).}
\label{fig:highlight_uas}
\end{figure}

The SWUF-3D UAS fleet, comprising 30 individual multicopters, was deployed during the sEOP in the IVTA. The drones were strategically distributed across the Nafingalm valley floor, with horizontal separation ranging from 10 to 500~m and flight heights up to 250~m above ground level (AGL). Over 100 fleet flights were executed during the daytime, complemented by nighttime vertical profiles up to 120~m. sIOP20 featured significant valley winds, forced by South Foehn over the ridge to the south of the measurement site. Figure~\ref{fig:highlight_uas}a shows wind speed and direction as well as color-coded turbulence kinetic energy (TKE) for one flight and all drones at 50~m (white contours) and 150~m~AGL (black contours). The data clearly reveal substantial flow variability, especially for TKE, even across these small scales. This successful measurement of turbulent flow across an Alpine valley using a dense network of in-situ instrumentation at such fine spatial scales is a first-time achievement for the TOC. A comprehensive understanding of the flow dynamics is envisioned through the synthesis of UAS data, concurrent Doppler wind lidar scans, and the established ground station network at Nafingalm (Section~\ref{sec:instrumentation}).
% FIGURE: UAS

Four MASC-MC systems were operated with a separation of 500~m along a valley transect (in the center of the valley, at the foot of the sidewall, at the slope and close to the crest) of the Inn Valley in the area of Radfeld during the sEOP (Fig.~\ref{fig:target_areas}). The four individual UAS performed about 400 vertical profiling flights up to 2~km~MSL, enabling the estimation of three-dimensional wind vectors, temperature, humidity and turbulence quantities, such as TKE, along the vertical profile. Figure~\ref{fig:highlight_uas}b shows an example of an up-valley wind situation during sIOP19. This case shows stronger up-valley winds  below the mountain ridge in the valley centre (yellow), with an easterly wind direction aligned with the valley axis, compared to the observation at the crest (red and purple). Due to vertical and horizontal wind shear, an increase in TKE is observed above the mountain ridge for the vertical profiles at the mountain crest in contrast to the valley centre. With the ability to observe turbulence variables simultaneously at multiple locations, it is possible to better understand the flow characteristics and dynamics of the ABL holistically, even in complex terrain. 

%% ---------- radiosoundings ----------
\subsection{Dual radiosoundings of gravity-wave breaking during the wEOP}
\label{subsec:techniques_radiosoundings}

Mountainous orography can significantly impact the atmospheric circulation by exerting a drag force on the background flow. This includes both (i) surface-level flow blocking and form drag and (ii) deep propagating gravity waves, which can transport and deposit horizontal momentum vertically to high altitudes. As weather and climate model resolutions increase, both the resolved and sub-grid parts of these drag forces must be \textit{scale aware} and change correctly with the grid spacing. Further, it is also essential that gravity wave momentum deposition is accurately simulated, especially in regions of high vertical wind shear that can often occur over mountains. Errors in these aspects, which are common in models, can lead to misrepresentations of the gravity wave drag on the background flow, resulting in inaccurate circulations in models.

During the TOC, a rarely used method of using dual radiosondes, launched simultaneously with a small horizontal separation, was used to measure the fine-scale horizontal structure of gravity waves and evaluate their representation in numerical models. The phase difference between two simultaneous launches, combined with the radiosondes' fine vertical sampling ($\approx$10~m), yields direct information about the physical scales of gravity waves. Paired with directional wind shear measurements (also observed by the radiosonde), scale-dependent gravity wave filtering and momentum deposition with height can be quantified. Wavelet analysis identifies co-varying waves in both profiles, which, across multiple IOPs, will give a scale-dependent view of gravity wave drag under different synoptic conditions. Such calculations are not possible with standard radiosonde launches, and will provide a robust test of existing techniques to validate the horizontal scales from single profiles. \citet{Shutts1994} tested this method for triple-launch cases over the Welsh mountains, where fluctuations in the radiosondes' rate of ascent were used to estimate the phase-line tilt with height. Since then, very few field campaigns have used this technique, and none in a highly mountainous region such as the Alps. This research will inform the development of scale-aware models.  

\begin{figure}
\centering
\includegraphics[width=\textwidth]{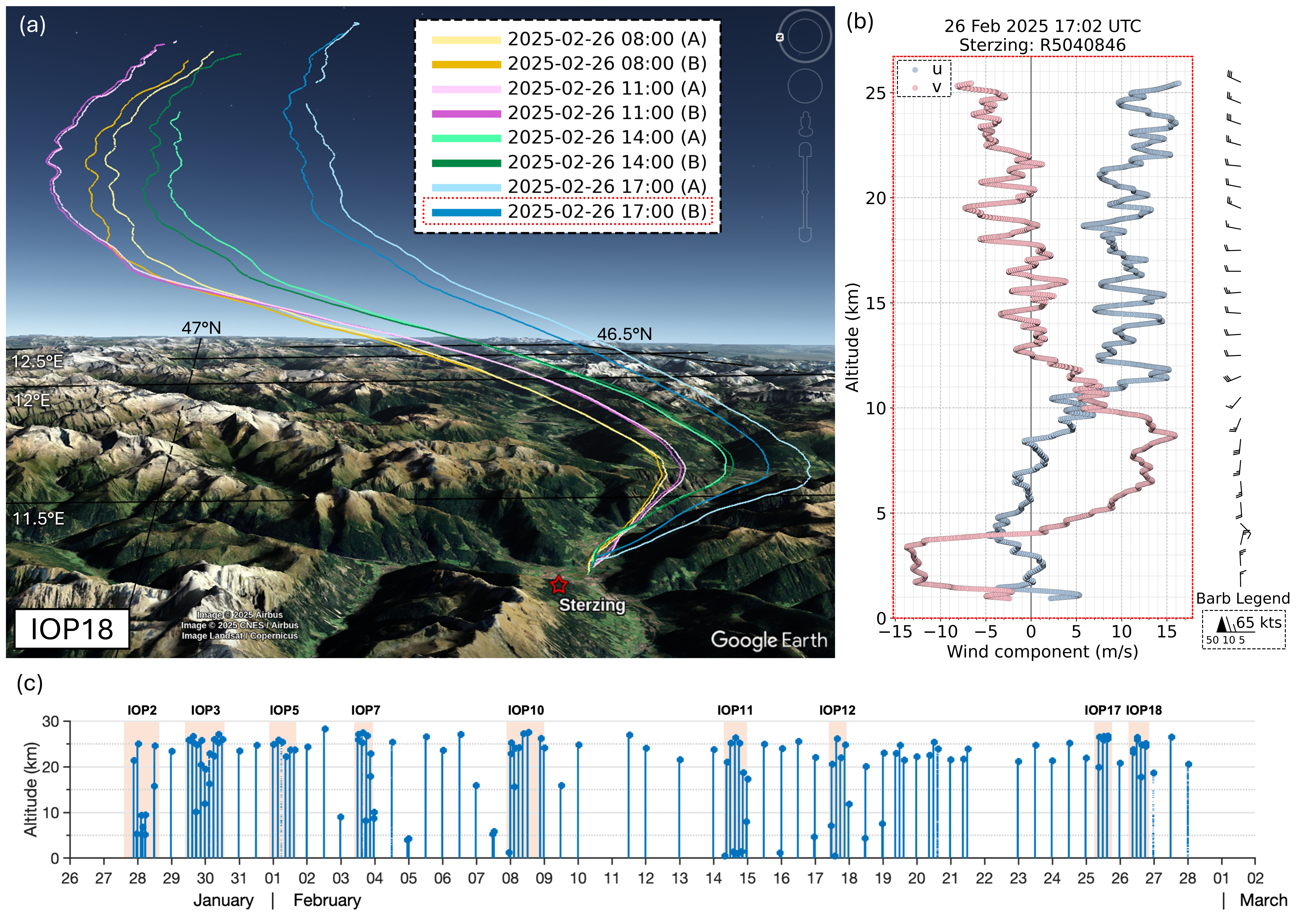}
\caption{Radiosonde launches from Sterzing (red star) during the wEOP, illustrated as (a) a 3D Google Earth view of the radiosonde tracks for wIOP18 (26~February 2025) (b) a wind-component profile of the 1700~UTC radiosonde launch during wIOP18 and (c) a timeseries showing all launches and their respective maximum altitudes. Blue and red profiles in (b) show the zonal ($u$) and meridional ($v$) wind components respectively, with wind barbs indicating wind speed and direction.}
\label{fig:highlight_dual_raso}
\end{figure}

During the wEOP, 9 gravity-wave IOPs were conducted (Tables~\ref{tab:iops} and \ref{tab:iops_weop}). Despite occasional technical difficulties prohibiting some dual launches, numerous dual radiosondes were launched at a 3-h cadence for a variety of different synoptic conditions, including deep propagating gravity waves, transient critical layers, and strong vertical wind shear. wIOP18 (Fig. \ref{fig:highlight_dual_raso}) is a good example of a complex synoptic situation in which the atmospheric environment exhibits both high directional wind shear and clear propagation of orographic gravity waves into the stratosphere, with four dual radiosonde launches during the IOP. Focusing on the 1700~UTC launch (Fig.~\ref{fig:highlight_dual_raso}b), low-level northerly winds below an altitude of 4~km rapidly reversed to become southerly just above the mountain peaks, before veering westerly as the radiosonde enters the stratosphere. Gravity-wave perturbations are clearly visible in the stratosphere (above $\approx$10~km altitude), with pronounced oscillations in both zonal and meridional wind components of $\approx$5~m~s$^{-1}$.  Figure~\ref{fig:highlight_dual_raso}a shows strong coherence between radiosonde pairs, and it is likely that some gravity wave filtering occurs as a result of the directional wind shear both at 4~km and near the tropopause, providing an ideal test case for high resolution models. This and other GW IOPs shown on the timeline in Fig~\ref{fig:highlight_dual_raso}c will be the focus of future work.
% FIGURE: dual radiosonde launches

%% ---------- model ----------
\subsection{High-resolution ICON runs for forecasting during the EOPs}
\label{subsec:techniques_icon}

To support experiment planning during the EOPs, the German Weather Service (DWD) provided dedicated high-resolution ICON \citep[Icosahedral Nonhydrostatic;][]{Zaengl2015,Zaengl2022} forecasts at a mesh size of 500~m for a large part of the Alpine domain. The forecasts were provided eight times per day using a model configuration named ICON-A05 (Alpine domain, 0.5 km). The configuration is based on the convection-permitting ICON-D2 model (2~km grid spacing for Germany and adjacent regions) and uses ICON's two-way nesting functionality to spawn higher-resolution, nested domains during the runtime of the forecasts. The ICON-A05 forecasts are thus initiated from the operational deterministic ICON-D2 analyses, with the nested domains being started after the latent-heat nudging phase at +40~min and +50~min for the 1-km and 500-m nests, respectively. An identical model configuration with nested domains over Germany (ICON-D05) has been running for operational forecasting at DWD since February 2025, and the model development work described in the following has been conducted for ICON-A05 and ICON-D05 in parallel.

To fully exploit the potential of the high model resolution, some model options not used in ICON-D2 were activated for the 500-m configurations. These include a 3D Smagorinsky closure for the horizontal (numerical) diffusion scheme, which is needed to impose enough mixing along the edges of strong convective updrafts, and a slope correction for the solar surface radiation including topographic shading. New developments for the 500-m configurations include a modified gust parameterization and a set of tuning modifications to improve the prediction of nocturnal surface inversions/cold-air pools in narrow valleys with steep slopes. Common gust parameterizations assume positive boundary-layer shear driven by surface friction, and that turbulent eddies are fully parameterized. The former assumption is violated for small-scale vertically decaying orographic gravity waves, which have (in the frictionless limit) their wind maximum right at the surface, whereas the latter assumption may not be fulfilled in deep convectively mixed boundary layers. The modified gust parameterization takes 10-min average 10-m winds rather than instantaneous values as input, and the excess gust speed is limited by the resolved wind maximum in the lowest 1500 m AGL times a resolution-dependent tuning factor. A detailed evaluation revealed that the limitation also benefits some phenomena not related to orography, for instance convective gust fronts. The modifications to improve nocturnal temperatures in valleys include tuning of the surface transfer scheme over steep slopes and of the TKE source term passed by the sub-grid scale orography scheme.

\begin{figure}
\centering
\includegraphics[width=\textwidth]{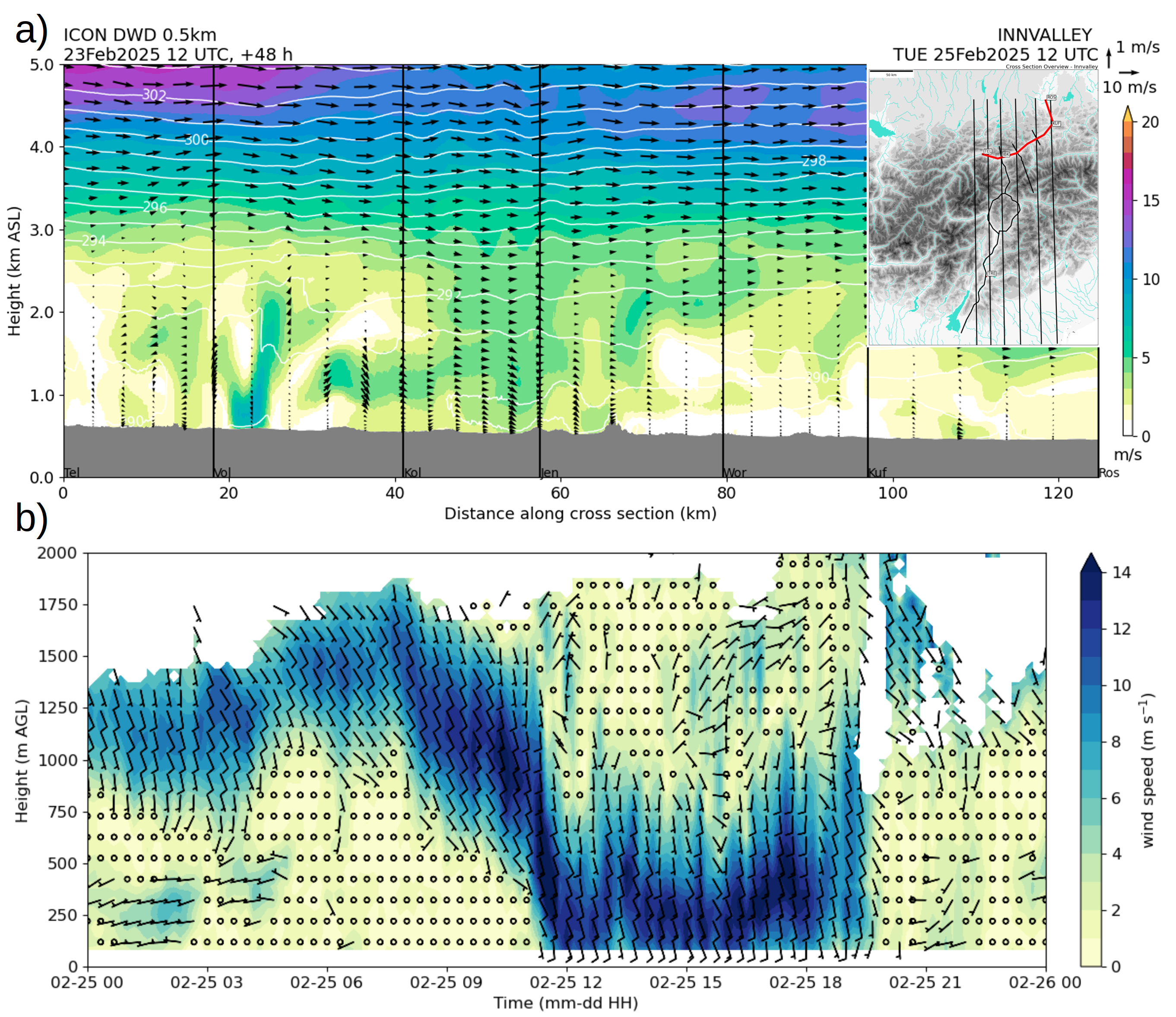}
\caption{(a) ICON-A05 forecast of horizontal wind speed and potential temperature along the Inn Valley for wIOP17 (25~February 2025, 1200~UTC). Small panel in the top-right shows the observational domain with the shown cross section highlighted in red. (b) Observed wind speed and direction from a Doppler wind lidar at Innsbruck during wIOP17.}
\label{fig:highlight_icon}
\end{figure}

As an example for the ICON-A05 performance, Fig.~\ref{fig:highlight_icon}a shows a model forecast of a shallow south foehn breakthrough in Innsbruck (at kilometre~22). Visible in the along-valley cross section are two key features of foehn events: a local maximum of the horizontal wind speed close to the surface and a nearly dry-adiabatic lapse rate up to an altitude of 2~km~MSL. With a lead time of 48~h, the ICON-A05 prediction of the time of foehn breakthrough at the valley floor is accurate to the hour, which occurs around 1100~UTC on 25 February (Fig.~\ref{fig:highlight_icon}b). Existing models and human forecasters still struggle to predict the occurrence and exact timing of these small-scale and local events at comparable lead times and with such accuracy.
% FIGURE: icon forecast

%% ---------- SCIENCE HIGHLIGHTS ----------
\section{First highlights---MoBL processes}
\label{sec:highlights}

%% ---------- mobl ----------
\subsection{MoBL structure and turbulence}
\label{subsec:highlight_mobl}

The MoBL is characterized by large spatial heterogeneity as a result of varying terrain and surface conditions and the influence of mountain-induced flow processes (e.g., thermally driven winds) across multiple scales \citep{Lehner2021}. During the wEOP and the sEOP, a network of EC stations was deployed in the IVTA on a steep, north-facing slope (Hochh\"auser) and on the floor of the Inn Valley, respectively, together with auxiliary instrumentation. The networks were designed to capture the local small-scale heterogeneity and its impact on advection in budgets of energy, momentum, and TKE.

\begin{figure}
\centering
\includegraphics[width=\textwidth]{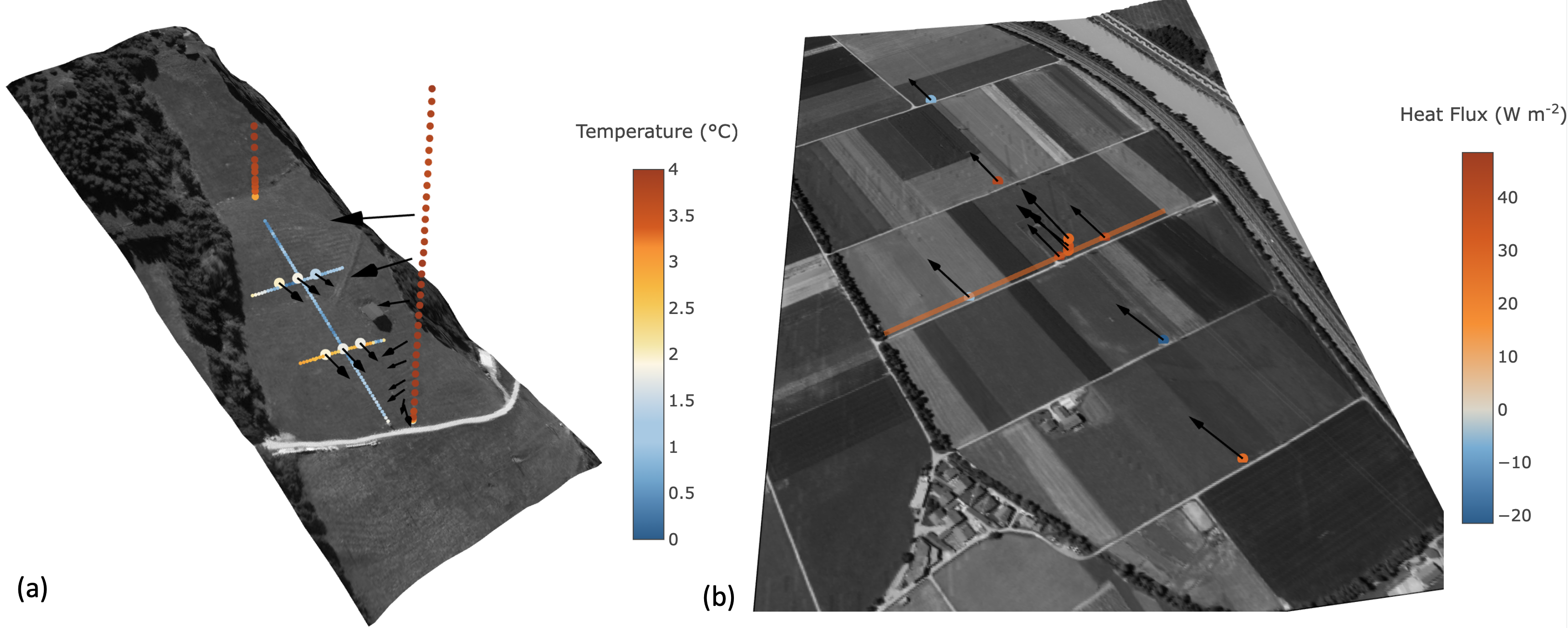}
\caption{Spatially distributed measurements on (a) a steep north-facing slope (IVTA) during wIOP9 (4~February 2025, 1600~UTC, 5-min averages) and (b) on a valley floor (IVTA) during sIOP3 (24~June 2025, 1430~UTC, 30-min averages) (a) Near-surface temperature from DTS observations, vertical temperature profile from a tethered-balloon sounding, and wind arrows from 2D sonic anemometers and a Doppler wind lidar, (b) heat flux from sonic anemometers (individual color dots) and a scintillometer (color line), wind arrows from 2D sonic anemometers. Sensible heat flux data from the EC stations are processed as in \citet{Stiperski2018}.}
\label{fig:highlight_ec_networks}
\end{figure}

The north-facing slope was mostly snow covered during the wEOP and received no direct solar radiation so that katabatic flows occurred regularly during undisturbed conditions. The early-evening example shown in Fig.~\ref{fig:highlight_ec_networks}a is from the almost 48-h long wIOP9 with clear-sky conditions leading to the formation of undisturbed katabatic flows. Both the katabatic flow and the temperature inversion above the slope remained, however, shallow, with a depth of about 10~m at 1600~UTC, transitioning to a westerly down-valley flow above and a close-to isothermal stratification. Near-surface temperatures from both a distributed temperature sensing (DTS) system and the station network highlight the spatial heterogeneity both along and across the slope axis, with distinctly higher temperatures in the eastern part of the slope and along the lower transect. Weaker variations on an even smaller spatial scale are also visible in the along-valley DTS transect.
% FIGURE: spatial variability - ec networks

The EC network during the sEOP was deployed across an area of approximately 500 $\times$ 1000~m$^{2}$, that is, the scale of a grid cell in a high-resolution operational model simulation. The afternoon example from a MoBL-T IOP in Fig.~\ref{fig:highlight_ec_networks}b  shows a well-developed easterly up-valley flow across the station network, with small variations in wind speed and direction. The sensible heat flux, however, reveals pronounced differences (data processing as in \citet{Stiperski2018}). The selected time is close to the afternoon transition when the magnitude of the heat flux is small and it changes from positive to negative values, which occurs typically relatively early in the afternoon at Kolsass \citep{Lehner2021}. Three of the stations have already changed sign at 1430~UTC or are close to 0~W~m$^{-2}$, with the others still positive. The three sites with a negative heat flux are, however, distributed across the network, thus not indicating a propagation of the transition from one direction or an obvious link to local surface properties. The magnitude of the sensible heat flux derived from the scintillometer agrees well with the heat flux from the sonic anemometers.

%% ---------- slope winds ----------
\subsection{Slope winds}
\label{subsec:highlight_slope_winds}

\begin{figure}
\centering
\includegraphics[width=\textwidth]{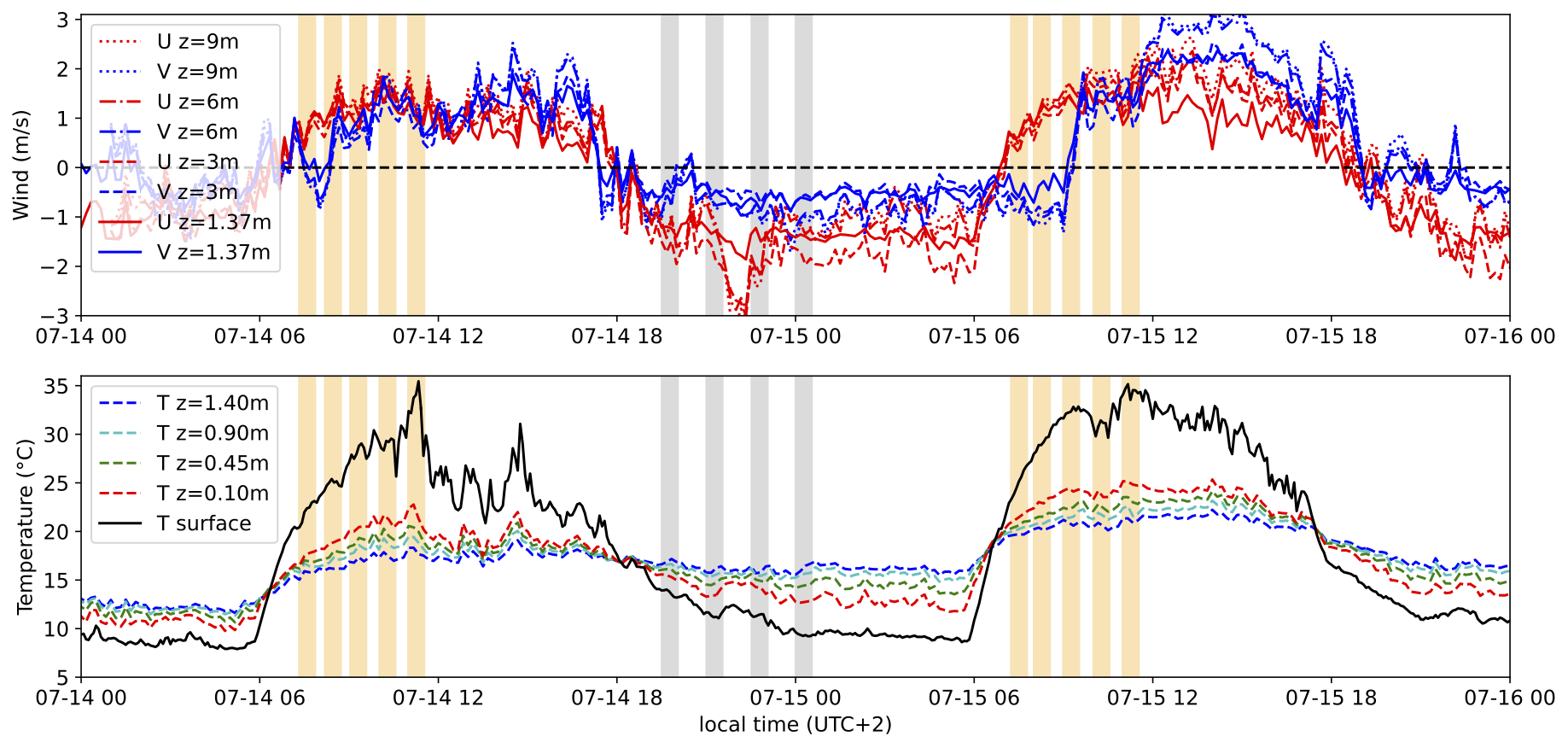}
\caption{48-hour time evolution (14--16~July 2025) of the 10-minute-averaged wind (top) and temperature (bottom), with the orange and grey bands highlighting the profiling periods. (Top) along-slope ($U$, red lines) and cross-slope ($V$, blue lines) wind components measured by 3D sonic anemometers at different heights. Positive velocities are defined upslope (i.e., easterly) for the along-slope component and up-valley (i.e., southerly) for the cross-slope component. (Bottom) Air temperature measured by thermocouples at different heights (dashed lines) and surface temperature derived from longwave radiation (solid black line).}
\label{fig:highlight_slope_winds}
\end{figure}

From mid-June to mid-October 2025 an intensive slope-wind campaign was conducted on Monte Baldo, a north--south oriented mountain range in the Italian pre-Alps between the Adige Valley and Lake Garda (AVTA, Fig.~\ref{fig:target_areas}). The main goal was to characterize the turbulence structure, transport processes, and surface-energy budget associated with thermally driven slope winds. The study site, located on a steep ($\approx 25^{\circ}$), east-facing slope covered by grass and low shrubs, was identified as well-suited for slope-wind research after a 2024 pre-campaign \citep{Rajput2026}. Building on this, the 2025 campaign employed multiple instruments and an observational setup designed to capture slope effects across spatial scales. The instruments included 2D and 3D sonic anemometers mounted on a three-level flux tower (3, 6, and 9~m~AGL) coupled with a two-level mast (1.37 and 2.80~m~AGL), wind and Raman lidars, and a profiler of near-surface wind and temperature. The latter consists of a vertically moving system with two multi-hole pitot tubes and thermocouples, covering heights from 2~mm to 2~m~AGL. Profiles were taken every hour at the early stage of anabatic wind onset and throughout the night for katabatic winds (orange and grey vertical bands in Fig.~\ref{fig:highlight_slope_winds}, respectively), with a whole vertical profile requiring 30~min. A similar profiler was also used at Hochhäuser during the wEOP (Section~\ref{subsec:highlight_mobl}) to determine near-surface turbulent properties. 
% FIGURE: slope winds

Figure~\ref{fig:highlight_slope_winds} shows the time evolution of wind and temperature at the flux tower and mast during two selected days, namely 14--15~July 2025. The weather was characterised by anticyclonic conditions and was thus favorable for thermally-driven flows. The diurnal cycles of wind clearly show the typical reversal from nighttime katabatic to daytime anabatic wind (Fig.~\ref{fig:highlight_slope_winds}, top). Downslope winds are observed at all levels, with maximum along-slope speed at 1--3~m~AGL. The slope wind persists steadily throughout the nighttime, apart from a wind gust observed around 2100~CET (UTC+2~h) on 14 July. Upslope winds are observed in the early morning (0700--1000~CET) before stronger lateral components appear (1000--1200~CET), followed by more complex convection episodes in the afternoon and interferences from larger-scale valley-driven circulations. The diurnal cycle of temperature (Fig.~\ref{fig:highlight_slope_winds}, bottom) shows a strong inversion during the nighttime katabatic flow (1800--0700~CET) between the ground and the adjacent approximately 1-m deep layer of the atmosphere.

%% ---------- convection ----------
\subsection{Convection and convective initiation}
\label{subsec:highlight_convection}

Recent research has shown that NWP models struggle to characterize the variability of precipitation with elevation \citep{Matiu2024} and to properly reproduce storms over orography \citep{Poujol2025}. \citet{Manzato2022} demonstrated that the Sarntal Alps (ACTA, Fig.~\ref{fig:target_areas}) record a peak in lighting occurrence in summer, suggesting favorable conditions for convective initiation. During the sEOP (May to mid-September), various institutions contributed to equipping this region with a suite of instruments to capture the initiation of convection and to quantify precipitation at different elevations. 

\begin{figure}
\centering
\includegraphics[width=\textwidth]{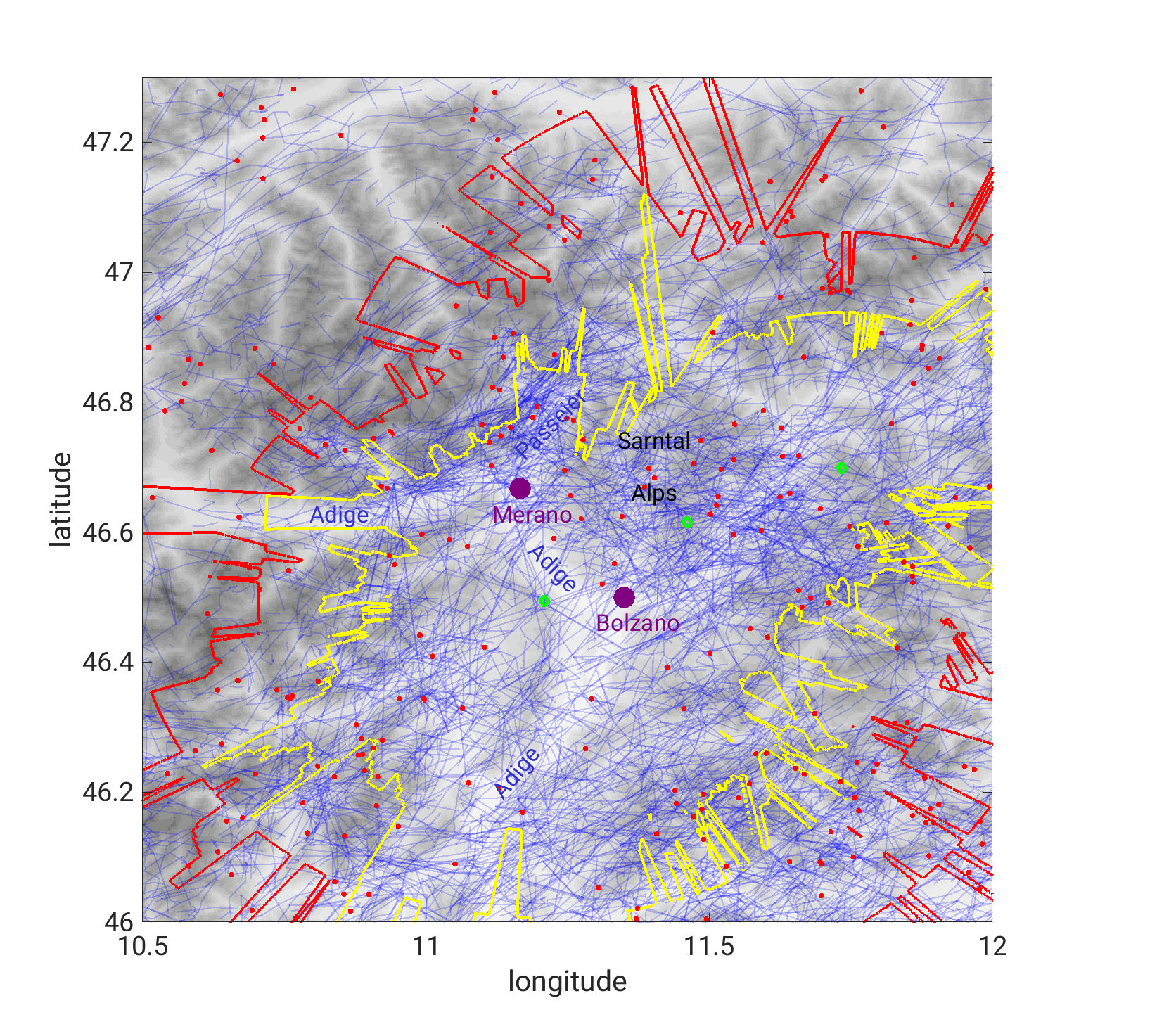}
\caption{Spatial distribution of observed thunderstorm tracks around the Sarntal Alps during the sEOP. The thin blue lines indicate the tracks of RCs. The red dots mark primary initiations. The green markers indicate the positions of the three radars: Monte Macaion, Rittner Horn, and Plose (from west to east). The yellow and red curves surround the area where at least one radar is able to observe down to 3000~m and 4000~m above sea level, respectively.}
\label{fig:highlight_thunderstorms}
\end{figure}

To detect when and where convective systems developed, measurements from three radars (volume scans with a 5-min temporal resolution) on Monte Macaion (C-Band, see Fig.~\ref{fig:highlight_thunderstorms} for the location), Rittner Horn (X-Band) and Plose (X-Band) are analyzed using the TRACE3D storm identification and tracking algorithm \citep{Handwerker2002}. TRACE3D identifies \textit{reflectivity cores} (RC) as continuous regions of strong echoes with a maximum reflectivity above 45~dBZ. These RCs are regarded as potential thunderstorms. A total of 17323 RCs were found in the region shown in Fig.~\ref{fig:highlight_thunderstorms} during the sEOP (an evolving thunderstorm is counted for each volume scan, i.e., every 5~min). Convective systems were generally observed after 1000~UTC, with a few occurrences even earlier. RC frequency peaks between 1700 and 1900~UTC, and by 2100~UTC the frequency has decreased to only a few thunderstorms. During the 39-day period of the sEOP, 24 days showed more than 100 RCs, 5 more than 1000 RCs (max. more than 2100 RCs on 30~June) and 11 days had no RCs at all. The spatial distribution of the RC tracks (Fig.~\ref{fig:highlight_thunderstorms}) reveals regions where RCs are identified frequently (e.g., along the Passeier Valley, above the Sarntal Alps) and rarely (especially in the Adige Valley between Bolzano and Merano). 
% FIGURE: thunderstorms

Thunderstorm initiation cannot be observed directly, only the first identification of an RC. We thus consider an RC a \textit{primary thunderstorm initiation} if no other RC has been identified within a radius of 30 km in the previous half hour. Otherwise, the new thunderstorm could have been caused by a predecessor. During the sEOP, 242 primary thunderstorm initiations occurred within the domain of Fig.~\ref{fig:highlight_thunderstorms} (red dots). From 0900 to 1900~UTC, ten or more primary initiations were detected per hour. Further analysis will explore the cross-links between meteorological condition (e.g. valley-wind strength, humidity, stability) and thunderstorm initiation, using multiple Doppler radar analyses as well as Doppler wind lidar and DIAL measurements at the KITcube stations around the Sarntal Alps.

\begin{figure}
\centering
\includegraphics[width=\textwidth]{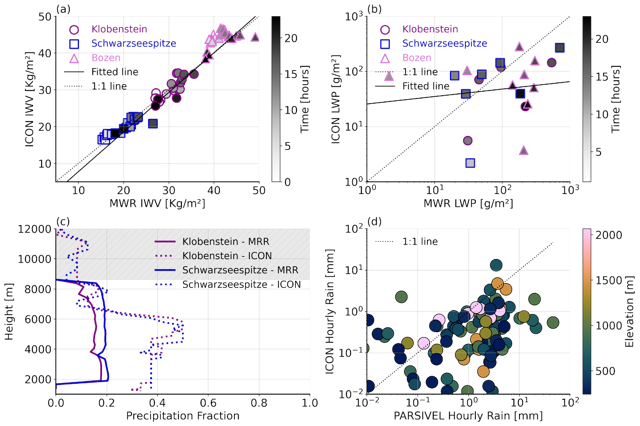}
\caption{Evaluation of ICON-A05 with observations from various sites in the Sarntal Alps for sIOP8. (a) Hourly mean integrated water vapor \citep[MWR retrieval][]{Marke2024} at Bozen (241~m~MSL), Klobenstein (1192~m~MSL), and Schwarzseespitze (2066~m~MSL) vs. ICON-A05 IWV in kg~m$^{-2}$ as a function of time of day (greyscale colorbar). (b) Same as (a) for LWP. (c) Precipitation fraction from micro rain radar and from ICON-A05 (sum of snow, graupel, and rain hydrometeors). The grey area indicates elevations that the radar signal cannot reach. (d) Cumulated hourly rain rate from a network of Parsivel disdrometers vs. ICON-A05 total cumulated rain as a function of elevation (colorbar). 1:1 lines are reported as black dotted lines, while linear fits are solid black lines.}
\label{fig:highlight_convection}
\end{figure}

The case study of sIOP8 (30 June; Fig.~\ref{fig:highlight_convection}) highlights the potential of ICON-A05 (Section~\ref{subsec:techniques_icon}) and the observational data to advance our knowledge of convection initiation in mountain environments and the subsequent precipitation processes. A westerly mesoscale flow initiated convective activity over the Alpine orography, with intense lightning activity over the entire mountain chain around noon, accompanied by deep convective clouds being advected eastward. The convective activity intensified and spread over the Po Valley during the afternoon, with precipitation recorded by all Parsivel disdrometers deployed in the Sarntal Alps from 1600 to 2200~UTC.
% FIGURE: convection

Figure~\ref{fig:highlight_convection}a shows that there is overall good agreement between ICON-A05 integrated water vapor (IWV) and observed values from microwave radiometers (MWRs) \citep[retrieval][]{Marke2024}. As expected, Bozen, the lowest site, displays the highest values, which increase from 40 to 50~kg~m$^{-2}$ during the day. ICON-A05 slightly overestimates IWV in the morning and underestimates it in the evening. IWV progressively decreases with elevation, with a stronger agreement between modelled and observed IWV at the two higher sites, without visible diurnal trends. Liquid water in clouds from ICON-A05 and observations \citep[retrieval][]{Marke2024} varies more strongly (Fig.~\ref{fig:highlight_convection}b) than IWV, which is probably mainly due to mismatches in the exact location of modeled and observed small clouds. Observed liquid water path (LWP) values between 100 and 400~g~m$^{-2}$ in the afternoon and evening at Bozen are, however, underestimated by the model by about a factor of two, whereas at higher elevations, LWP values mostly below 200~g~m$^{-2}$ are better captured. The model thus seems to correctly describe the water vapour and cloudiness on this day.

Fig.~\ref{fig:highlight_convection}c compares the daily precipitation fraction derived from the model with radar reflectivity profiles from vertically pointing K-band radars (Micro Rain Radar). The observed daily precipitation fraction is defined here as the ratio  between the number of radar bins with meaningful radar reflectivity and the total number of radar bins observed during the day for each range gate. The model precipitation fraction at each model level is the ratio between the number of model cells for which the cloud water and ice mixing ratios are below $10^{-6}$~kg~kg$^{-1}$ and at least one among the snow, rain, and graupel mixing ratios is larger than $10^{-6}$~kg~kg$^{-1}$ and the total number of model cells during the day. ICON-A05 overestimates the number of periods with rainfall at both sites. The observed profiles are largely constant with height up to 8000~m, while the modelled profiles are only constant between 2000 and 4000~m, with higher values between 4000 and 6000~m, before decreasing to zero aloft. Figure~\ref{fig:highlight_convection}d shows that despite the higher frequency, the hourly accumulated precipitation amounts at the ground are underestimated by ICON-A05. This underestimation agrees with other 500-m ICON evaluation studies of precipitation (Daniela Littman, personal communication, 11~Dec 2025).

This preliminary analysis of a single case study does not allow final conclusions about the model's potential biases, but highlights the capabilities of the collected datasets. The combination of hectometer-scale simulations and a unique observational dataset allow monitoring convective processes within the MoBL to tackle open research questions, including how convection is initiated over complex terrain and what is the variability of precipitation profiles and amounts with elevation.

%% ---------- SUMMARY AND OUTLOOK ----------
\section{Summary and outlook}
\label{sec:summary}

A one-year long measurement campaign was conducted in the European Alps as part of the international research programme TEAMx (TEAMx Observational Campaign, TOC). With TEAMx focusing on the transport and exchange processes in the atmosphere over complex, mountainous terrain, the main goal for the TOC was to collect a unique dataset that can be used to further our understanding of these processes and for model evaluation and development. In total, more than 40 institutions participated in the TOC (Table~\ref{tab:institutions}), in particular during the two 6-week long Extended Observational Periods (EOPs) in winter and summer 2025. A number of Intensive Observational Periods (IOPs) was conducted during both EOPs (Tables~\ref{tab:iops},\ref{tab:iops_weop},\ref{tab:iops_seop}), focusing on key processes contributing to the total transport and exchange of mass, energy, and momentum, including gravity waves, convection, the spatio-temporal evolution of the Mountain Boundary Layer (MoBL), and thermally driven slope winds. About 10 sites within the four TEAMx target areas (Fig.~\ref{fig:target_areas}) were heavily instrumented during these periods with Doppler wind lidars, temperature and humidity profilers, Raman lidars, ceilometers, radiosonde stations, UAS activities, eddy-covariance towers, and other instruments (Fig.~\ref{fig:instrumentation}). Additional spatial measurements were collected with three research aircraft (Fig.~\ref{fig:flight_tracks}).

A wealth of observational data was thus collected as part of the TOC. While the analysis of this rich dataset has barely started, first and preliminary results (e.g., Section~\ref{sec:highlights}) highlight its potential for both process understanding and model evaluation. The scope and international character of TEAMx and the TOC encourage collaborations across research groups, for example, by combining complementary datasets from different instruments for joint analysis of case studies. In addition, the collaboration of multiple research groups with expertise with similar instruments allows the development of common post-processing strategies and retrieval algorithms, for example, post-processing of eddy-covariance data, deriving turbulence statistics from Doppler wind lidars, and retrieval algorithms for microwave temperature and humidity profilers. Existing and new collaborations are being actively supported by bringing the TEAMx community together at regular intervals, with TEAMx Workshops planned to take place annually for the next years.

Experience from previous field campaigns, such as the Mesoscale Alpine Programme (MAP), suggests that such datasets are being actively used for more than a decade with the number of publications peaking about five years after the campaign \citep{Volkert2007}. We do of course hope that the TOC dataset will lead to a similar number of research studies and new findings. To facilitate the use of the collected data, both within and outside the current TEAMx research community, a TEAMx Data Management Plan has been written that outlines the guidelines for using and sharing the data. All TOC datasets will thus be published one year after the end of the TOC in different public data repositories (e.g., zenodo or PANGAEA), with all public datasets linked and thus easily findable through a joint TEAMx entry on the Earth Data Portal (https://earth-data.de/collections/46), a data hub of seven research centers of the Helmholtz Association in the fields of earth and environment.

The high-resolution forecast runs provided by several European weather services during the TOC (Table~\ref{tab:nwp}) provided not only an important, but also a unique, contribution to the TOC. They proved to be invaluable tools for IOP decisions during both EOPs, for example, aiding in predicting of low-level stratiform clouds in the Inn Valley during the wEOP which delayed flight operations in the mornings and in predicting the timing of thermally driven flows and convection during the sEOP. Many of these forecast runs have also been archived and will be used for the analysis of case studies, model intercomparison studies, and model evaluation.

The collection of the TOC dataset is only one of four TEAMx main goals. This dataset is, however, a significant building block for the other three goals, that is, (i) improving our understanding of the transport and exchange processes in the atmosphere over mountains, (ii) improving the representation of these processes in numerical weather and climate models, and (iii) improving impact models (e.g., air-pollution transport and hydrological models) by sharing results and data with weather and climate service providers. With one goal of TEAMx fully accomplished as soon as the data are published, focus now shifts to the analysis and numerical modeling to exploit this rich dataset. In particular, numerical modeling studies will provide key contributions in the near future, including both idealized and real-case studies, mesoscale to large-eddy simulations, and a high-resolution re-analysis for part of the TOC.

%% ---------- ACKNOWLEDGMENTS -------
\section*{Acknowledgments}
% funding grants
Funding: This work was supported by the Austrian Science Fund FWF (grant number 10.55776/PIN2143424 - M. Lehner), the German Federal Ministry for Transport BMV through the IDEA-S4Snetwork (grant number 4823IDEAP5 - C. Acquistapace, D. Corradini; grant number 4823IDEAP6 - A. Oertel), the ``Rita Levi Montalcini'' young researcher program (Code PGR217DXNP - C. Acquistapace), the Natural Environment Research Council NERC (grant number NE/Z50399X/1 - T. Banyard, N. Hindley, A. Orr, A. Ross, S. Gumber; grant number NE/Z503964/1 - H. Dacre, N. Harvey, N. Sharma; grant number NE/X017842/1 - N. Hindley; grant number NE/W003201/1 - N. Hindley; grant number NE/Z503988/1 - I. Renfrew, J. Mustafa, D. Smith; grant number NE/Z503952/1 - A. Ross), the French National programme LEFE (Les Enveloppes Fluides et l’Environnement) 2024 (C. Brun), the Labex OSUG@2020 Investissements d’avenir ANR10 LABX56 (C. Brun), the European Union under the Next Generation EU PRIN 2022 Prot. n. 2022NEWP4J (CUP B53D23007350006 - W. Cairns; CUP E53D23004450006 - D. Zardi), the Deutsche Forschungsgemeinschaft (German Research Foundation) DFG (grant number 563995063 - P. Gasch; grant number 549349557 - A. Platis, M. Kippenberger), the Euregio Science Fund (grant number IPN 187 - L. Giovannini, N. Vendrame), the European Research Council (ERC) under the European Union's Horizon 2020 research and innovation program (grant agreement no. 101001691 - I. Stiperski; grant agreement no. 101040823 - N. Wildmann, A. Alexa), ACTRIS-D funded by the Federal Ministry of Research, Technology and Space (grant number 01LK2001B - H. Vogelmann; grant number 01LK2001B - J. V\"ullers), the Helmholtz Research Program Changing Earth – Sustaining our Future within the Helmholtz research field Earth and Environment (H. Vogelmann), the Italian Space Agency (ASI) and Ministry of University and Research (MUR) (Contract No. 2024-5-E.0 - CUP No. I53D24000060005 - D. Zardi), the consortium “iNEST” (Interconnected Nord-Est Innovation Ecosystem) funded by the European Union under NextGenerationEU (Mission 4.2, Investment 1.5, project no. ECS 00000043 - D. Zardi), the CSP LINKS Foundation Torino-POC-Transition (n. 121984 - S. Abdunabiev, D. Tordella) the Swiss National Science Foundation (SNSF, grant number 200020-212121 - F. Conen, A. Einbock), the US National Science Foundation NSF (grant number ATM-2332467 - S. de Wekker), the University of Innsbruck and GeoSphere Austria through the Innsbruck Network for Weather and Climate Research IWCR (D. Deacu), the Austrian Research Promotion Agency FFG (grant number FO999915175 - M. Dorninger), the Research Council of Norway RCN (grant number 302458 - K. Haualand; grant number 227777 - S. Kral, J. Reuder), the French National Research Agency ANR (grant number 22-PETA-0008 - S. Sch\"afer, J. Barri\'e, N. Perche, O. Garrouste, J.-C. Etienne, Q. Libois).

% individuals
The authors would like to thank all the students, technicians, and volunteers who helped to set up and maintain the instruments and data transfer in the field, those who provided support with initial data processing, the many landowners who allowed us to deploy equipment on their land, and everyone else who supported the TOC in whichever capacity.

%% ---------- APPENDIX ----------
\appendix

% --- co-authors 'TOC team'
\newpage
\section{TOC Team}
\label{app:coauthors}
TOC team members listed as co-authors at the beginning of the paper are not repeated in the list below.
\begin{scriptsize}
\begin{longtable}{p{3.5cm} p{9.5cm}}
    Shahbozbek Abdunabiev & Politecnico di Torino, Department of Applied Science and Technology \\
    Almut Alexa & Deutsches Zentrum f\"ur Luft- und Raumfahrt e.V., Institut f\"ur Physic der Atmosph\"are; University of Innsbruck, Department of Atmospheric and Cryospheric Sciences \\
    Doug Anderson & FAAM Airborne Laboratory \\
    Jens Bange & Eberhard Karls University of Tuebingen, Department of Geosciences \\
    Jo\"el Barri\'e & M\'et\'eo-France CNRS, University of Toulouse CNRM \\
    S\'ebastien Blein & M\'et\'eo-France CNRS, University of Toulouse CNRM \\
    Alessandro Bracci & National Research Council, Institute of Atmospheric Sciences and Climate (CNR-ISAC) \\
    Sebastiano Carpentari & University of Trento, Department of Civil, Environmental and Mechanical Engineering \\
    Massimo Cassiani & University of Trento, Department of Civil, Environmental and Mechanical Engineering; Norwegian Institute for Air Research \\
    Franz Conen & University of Basel, Department of Environmental Sciences \\
    Daniele Corradini & University of Cologne, Institute of Geophysics and Meteorology \\
    Federico Dallo & National Research Council, Institute of Polar Sciences (CNR-ISP) \\
    Amitraj Dash & Met Office \\
    Stephan F. J. de Wekker & University of Virginia, Department of Environmental Sciences \\
    Daniel Deacu & GeoSphere Austria, Regionalstelle Tirol und Vorarlberg \\
    Marco Di Paolantonio & National Research Council, Institute of Marine Sciences (CNR-ISMAR) \\
    Manfred Dorninger & University of Vienna, Department of Meteorology and Geophysics \\
    Andreas D\"ornbrack & Deutsches Zentrum f\"ur Luft- und Raumfahrt e.V., Institut f\"ur Physic der Atmosph\"are \\
    Viktoria D\"urlich & Karlsruhe Institute of Technology, Institute of Meteorology and Climate Research Troposphere Research \\
    Annika Einbock & University of Basel, Department of Environmental Sciences \\
    Anselm Erdmann & Karlsruhe Institute of Technology, Institute of Meteorology and Climate Research Troposphere Research \\
    Jean-Claude Etienne & M\'et\'eo-France CNRS, University of Toulouse CNRM \\
    Alex Fearn & National Centre for Atmospheric Science; University of Leeds, School of Earth, Environment and Sustainability \\
    Thomas Feuerle & TU Braunschweig, Institute of Flight Guidance \\
    Angelo Finco & Catholic University of the Sacred Heart, Department of Mathematics and Physics \\
    Andreas Friesinger & University of Innsbruck, Department of Atmospheric and Cryospheric Sciences \\
    Olivier Garrouste & M\'et\'eo-France CNRS, University of Toulouse CNRM \\
    Mauro Ghirardelli & University of Innsbruck, Department of Atmospheric and Cryospheric Sciences \\
    Brigitta Goger & ETH Zurich, Institute for Atmospheric and Climate Science; present affiliation: GeoSphere Austria, Analysis and Model Development \\
    Rebecca Gugerli & Federal Office of Meteorology and Climatology MeteoSwiss \\
    Siddharth Gumber & British Antarctic Survey \\
    Natalie Harvey & University of Reading, Department of Meteorology \\
    Kristine Flack\'e Haualand & Western Norway University of Applied Sciences, Department of Civil Engineering and Environmental Sciences \\
    Maxime Hervo & Federal Office of Meteorology and Climatology MeteoSwiss \\
    Werner Jud & University of Innsbruck, Department of Atmospheric and Cryospheric Sciences \\
    Rudolf Kaltenb\"ock & Austro Control \\
    Moritz Kippenberger & Eberhard Karls University of Tuebingen, Department of Geosciences \\
    Stephan T. Kral & University of Bergen, Geophysical Institute; Bjerknes Centre for Climate Research \\
    Larissa Lacher & Karlruhe Institute of Technology, Institute of Meteorology and Climate Research Atmospheric Aerosol Research \\
    Francesca Lappin & Deutsches Zentrum f\"ur Luft- und Raumfahrt e.V., Institut f\"ur Physik der Atmosph\"are \\
    Hao Li & Karlruhe Institute of Technology, Institute of Meteorology and Climate Research Atmospheric Aerosol Research \\
    Quentin Libois & M\'et\'eo-France CNRS, University of Toulouse CNRM \\
    Christian Mallaun & Deutsches Zentrum f\"ur Luft- und Raumfahrt e.V., Einrichtung Flugexperimente \\
    Riccardo Marzuoli & Catholic University of the Sacred Heart, Department of Mathematics and Physics \\
    Herv\'e Michallet & Universit\'e Grenoble Alpes, Laboratoire des \'Ecoulements G\'eophysiques et Industriels \\
    Mario Marcello Miglietta & National Research Council, Institute of Atmospheric Sciences and Climate (CNR-ISAC) \\
    Ottmar M\"ohler & Karlruhe Institute of Technology, Institute of Meteorology and Climate Research Atmospheric Aerosol Research \\
    Stefan M\"uller & University of Innsbruck, Department of Atmospheric and Cryospheric Sciences \\
    Jack M. Mustafa & University of East Anglia, School of Environmental Sciences; University of Leeds, School of Earth, Environment and Sustainability \\
    Vladyslav Nenakhov & Deutsches Zentrum f\"ur Luft- und Raumfahrt e.V., Einrichtung Flugexperimente \\
    Ulrike Nickus & University of Innsbruck, Department of Atmospheric and Cryospheric Sciences \\
    Annika Oertel & Karlsruhe Institute of Technology, Institute of Meteorology and Climate Research Troposphere Research \\
    Sarah Paratoni & Karlsruhe Institute of Technology, Institute of Meteorology and Climate Research Troposphere Research; present affiliation: University of Innsbruck, Department of Atmospheric and Cryospheric Sciences \\
    Ferdinando Pasqualini & National Research Council, Institute of Atmospheric Sciences and Climate (CNR-ISAC) \\
    Noah Perche & M\'et\'eo-France CNRS, University of Toulouse CNRM \\
    Lena Pfister & University of Innsbruck, Department of Atmospheric and Cryospheric Sciences, present affiliation: Stadt Innsbruck, Referat Stadtklima und Umwelt \\
    Nathan Philippot & Universit\'e Grenoble Alpes, Laboratoire des \'Ecoulements G\'eophysiques et Industriels \\
    Davide Plebani & Catholic University of the Sacred Heart, Department of Mathematics and Physics; Katholeike Universiteit Leuven, Department of Earth and Environmental Sciences \\
    Davide Poggi & Politecnico di Torino, Department of Environment, Land and Infrastructure Engineering \\
    Bernhard Pospichal & University of Cologne, Institute of Geophysics and Meteorology \\
    Michal Posyniak & Karlsruhe Institute of Technology, Institute of Meteorology and Climate Research Atmospheric Environmental Research \\
    Rainer Prinz & University of Innsbruck, Department of Atmospheric and Cryospheric Sciences \\
    Joachim Reuder & University of Bergen, Geophysical Institute; Bjerknes Centre for Climate Research \\
    Lindsay Rhodes & National Centre for Atmospheric Science; University of Leeds, School of Earth, Environment and Sustainability \\
    Phil Rosenberg & National Centre for Atmospheric Science; University of Leeds, School of Earth, Environment and Sustainability \\
    Claudia Rossetti & National Research Council, Institute of Polar Sciences (CNR-ISP) \\
    Beth Saunders & University of Innsbruck, Department of Atmospheric and Cryospheric Sciences \\
    Loren Schaeffler & University of Innsbruck, Department of Atmospheric and Cryospheric Sciences; present affiliation: Karlsruhe Institute of Technology, Institute of Meteorology and Climate Research Troposphere Research \\
    Sophia Sch\"afer & M\'et\'eo-France CNRS, University of Toulouse CNRM \\
    Martin Sch\"on & Eberhard Karls University of Tuebingen, Department of Geosciences \\
    Katrin Sedlmeier & Deutscher Wetterdienst (DWD), Regionales Klimab\"uro M\"unchen \\
    Nischal Sharma & University of Reading, Department of Meteorology \\
    Xuefeng Shi & Karlruhe Institute of Technology, Institute of Meteorology and Climate Research Atmospheric Aerosol Research \\
    Daniel Smith & University of East Anglia, School of Environmental Sciences \\
    Johannes Speidel & Karlsruhe Institute of Technology, Institute of Meteorology and Climate Research Atmospheric Environmental Research \\
    Daniela Tordella & Politecnico di Torino, Department of Applied Science and Technology \\
    Johannes Vergeiner & GeoSphere Austria, Regionalstelle Tirol und Vorarlberg \\
    Hannes Vogelmann & Karlsruhe Institute of Technology, Institute of Meteorology and Climate Research Atmospheric Environmental Research \\
    Andrea Wiech & Deutsches Zentrum f\"ur Luft- und Raumfahrt e.V., Institut f\"ur Physic der Atmosph\"are; University of Innsbruck, Department of Atmospheric and Cryospheric Sciences \\
    Corwin Wright & University of Bath, Centre for Climate, Adaptation and Environment Research \\
\end{longtable}
\end{scriptsize}

% --- Participating institutions ---
\newpage
\section{Participating institutions}
\label{app:institutions}

\begin{footnotesize}
\begin{longtable}{p{8cm} p{1.5cm} p{2.5cm}}
    \hline
    Institution & Country & Target Area \\
    \hline
    Austro Control & AUT & IVTA \\
    Autonome Provinz Bozen--S\"udtirol--Amt f\"ur Meteorologie und Lawinenwarnung & ITA & ACTA \\
    BOKU University & AUT & IVTA \\
    British Antarctic Survey & GBR & ACTA \\
    Bundesamt f\"ur Meteorologie und Klimatologie MeteoSchweiz & CHE & IVTA/forecasts \\
    Consiglio Nazionale delle Ricerche--Istituto di Scienze dell'Atmosfera e del Clima & ITA & AVTA \\
    Consiglio Nazionale delle Ricerche--Istituto di Scienze Polari & ITA & AVTA \\
    DLR--Deutsches Zentrum f\"ur Luft- und Raumfahrt & GER & IVTA, ACTA \\
    DWD--Deutscher Wetterdienst & GER & IVTA, forecasts \\
    Eberhard Karls Universit\"at T\"ubingen & GER & IVTA \\
    GeoSphere Austria & AUT & IVTA/forecasts \\
    H{\o}gskulen p\r{a} Vestlandet & NOR & IVTA \\
    Karlsruher Institut f\"ur Technologie--Institut f\"ur Meteorologie und Klimaforschung Atmosph\"arische Aerosol Forschung & GER & ACTA \\
    Karlsruher Institut f\"ur Technologie--Institut f\"ur Meteorologie und Klimaforschung Atmosph\"arische Spurengase und Fernerkundung & GER & IVTA \\
    Karlsruher Institut f\"ur Technologie--Institut f\"ur Meteorologie und Klimaforschung Atmosph\"arische Umweltforschung & GER & PATA, IVTA \\
    Karlsruher Institut f\"ur Technologie--Institut f\"ur Meteorologie und Klimaforschung Troposph\"arenforschung & GER & ACTA \\
    Land Tirol--Abteilung Waldschutz & AUT & IVTA \\
    Ludwig-Maximilians-Universit\"at M\"unchen & GER & PATA \\
    Met Office & GBR & forecasts \\
    Météo-France & FRA & IVTA/forecasts \\
    National Centre for Atmospheric Science & GBR & AVTA, ACTA, IVTA, PATA \\
    NILU & NOR & AVTA \\
    Politecnico di Torino & ITA & ACTA \\
    Technische Universit\"at Braunschweig & GER & IVTA, ACTA \\
    Technische Universit\"at Wien & AUT & IVTA \\
    Umweltbundesamt & GER & PATA \\
    Umweltforschungsstation Schneefernerhaus & GER & PATA \\
    Universit\`a Cattolica di Brescia & ITA & AVTA \\
    Universit\`a degli Studi della Basilicata & ITA & AVTA \\
    Universit\`a degli Studi di Trento & ITA & AVTA \\
    Universit\`a di Bologna & ITA & AVTA \\
    Universit\"at Basel & CHE & AVTA \\
    Universit\"at Innsbruck & AUT & IVTA \\
    Universit\"at Wien & AUT & ACTA \\
    Universit\"at zu K\"oln & GER & ACTA \\
    Universit\'e Grenoble Alpes & FRA & IVTA, AVTA \\
    Universitetet i Bergen & NOR & IVTA \\
    University of Bath & GBR & ACTA \\
    University of East Anglia & GBR & IVTA, ACTA \\
    University of Leeds & GBR & ACTA \\
    University of Reading & GBR & IVTA, ACTA \\
    University of Virginia & USA & IVTA \\
    \hline
\caption{List of institutions participating in the TOC, including the target areas of their observational activities. Institutions are listed alphabetically in their respective national language.}
\label{tab:institutions}
\end{longtable}
\end{footnotesize}

% --- IOP overview ---
\newpage
\section{IOP overview}
\label{app:iops}

\begin{landscape}
\begin{longtable}{p{0.8cm} p{1.7cm} p{4cm} p{1cm} p{12cm}}
    \hline
    wIOP & Type    & Time (UTC) & Dual raso & Characterization \\
    \hline
    1 & SLP & 24 Jan 11--25 Jan 00 & - & weak foehn influence \\
    2 & GW & 27 Jan 14--28 Jan 12 & 4 & strong southerly synoptic winds; deep wave propagation \\
    3 & GW & 29 Jan 11--30 Jan 09 & 7 & weak synoptic winds; transient critical layer \\
    4 & SLP & 30 Jan 14--31 Jan 03 & - & interactions between katabatic flows, gravity waves, and synoptic winds; increasing cloud cover \\
    5 & GW & 31 Jan 23--1 Feb 15 & - & weak to moderate southerly synoptic winds; strong directional wind shear \\
    6 & SLP & 2 Feb 14--3 Feb 06 & - & katabatic flow conditions; low stratus in the Inn Valley \\
    7 & GW & 3 Feb 11--4 Feb 00 & 5 & weak to moderate gravity-wave activity, north-easterly flow \\
    8 & SLP & 3 Feb 14--4 Feb 07 & - & katabatic flows \\
    9 & SLP & 4 Feb 08--6 Feb 06 & - & clear skies \\
    10 & GW & 7 Feb 23–-9 Feb 00 & 2 & strong westerly to south-westerly synoptic winds; strong directional wind shear \\     
    11 & GW & 14 Feb 08--15 Feb 00 & - & moderately strong gravity waves; northerly flow; transient critical layer \\
    12 & GW & 17 Feb 11--21 & 1 & relatively strong north-westerly winds; deep wave propagation \\
    13 & SLP & 18 Feb 17-–19 Feb 09 & - & katabatic winds \\
    14 & MoBL-T & 19 Feb 05-–18 & - & low stratus in the Inn Valley dissolved before morning; moisture in low layers \\
    15 & MoBL-T & 20 Feb 05--18 & - & thermally driven conditions \\
    16 & MoBL-D & 21 Feb 05–-18 & - & low stratus in the Inn Valley; increasing dynamical influence \\
    17 & GW & 25 Feb 08--15 & 3 & short-lived gravity-wave event; moderately strong south-westerly winds; strong directional wind shear \\
    18 & GW & 26 Feb 08--18 & 4 & moderately strong westerly winds; strong directional wind shear \\
    \hline
\caption{IOPs during the wEOP, with their start and end time, the number of dual radiosonde launches at Sterzing (Section~\ref{subsec:techniques_radiosoundings}), and a preliminary characterization. IOP types are defined in Section~\ref{sec:toc}.}
\label{tab:iops_weop}
\end{longtable}
\end{landscape}

\begin{landscape}
\begin{longtable}{p{0.8cm} p{1.7cm} p{1.2cm} p{4cm} p{12cm}}
    \hline
    sIOP & Type & TA  & Time (UTC) & Characterization \\
    \hline
    1 & VEJ & IVTA & 21 Jun 17--22 Jun 11 & valley-exit jet with up to 12~m~s$^{-1}$ \\ 
    2 & SLP & AVTA & 22 Jun 05--23 Jun 10 & clear-sky, anticyclonic conditions; cumulus in the afternoon \\
    3 & MoBL–T & ACTA & 24 Jun 11--23 & well developed valley winds; shallow cumulus above ridges; \\
    4 & CON & ACTA & 25 Jun 08--23 & well developed valley winds; developing cumulus \\
    5 & SLP & AVTA & 27 Jun 02--30 Jun 08 & clear-sky conditions; north-foehn influence \\
    6 & MoBL-T & ACTA & 28 Jun 05--20 & weak valley winds; strong north-foehn influence; shallow cumulus above ridges \\
    7 & MoBL-T & IVTA & 29 Jun 02--23 & well-developed valley winds \\
    8 & CON & ACTA & 30 Jun 05–-20 & deep moist convection south of the main Alpine crest \\
    9 & CON & ACTA & 1 Jul 05--17 & weak valley winds; cumulus mediocris to congestus above ridges \\
    10 & MoBL-T & IVTA & 2 Jul 02--19 & mostly clear-sky; developing cumulus above main Alpine crest \\
    11 & CON & ACTA & 5 Jul 05--17 & widespread cumulus mediocris; cirrus from MCS dampened convective activity and valley-wind development \\
    12 & GW & IVTA, ACTA & 9 Jul 02--10 Jul 23 & north-westerly flow; moderate mountain-wave activity; broken cloud cover with occasional rain showers \\
    13 & MoBL-T & IVTA, ACTA & 11 Jul 02--17 & relatively dense cloud cover  \\
    14 & MoBL–T & ACTA & 12 Jul 05--17 & altostratus and cumulus \\
    15 & MoBL-T & IVTA & 13 Jul 02--17 & convective clouds and rain showers \\
    16 & MoBL-T & IVTA, ACTA & 15 Jul 05–-17 & dense cloud cover and rain showers \\
    17 & MoBL-T & ACTA & 16 Jul 05--17 & no valley-wind circulation; rain showers \\
    18a & VEJ & IVTA & 18 Jul 05--23 & valley-exit jet up to 13~m~s$^{-1}$ in the morning \\
    18b & MoBL–T & IVTA, ACTA & 18 Jul 05--23 & well developed valley winds; cumulus over north ridges \\
    19a & MoBL-T & IVTA & 19 Jul 02--23 & well developed valley winds; isolated shallow cumulus in the morning; increasing cloud cover and rain during the afternoon \\
    19a & CON & ACTA & 19 Jul 02--23 & isolated shallow cumulus in the morning; increasing cloud cover and rain during the afternoon \\
    20 & MoBL-T & IVTA & 20 Jul 02--23 & weak valley winds; weak gravity waves; shallow clouds \\
    21a & MoBL-T & IVTA, ACTA & 22 Jul 05--23 & well developed valley winds; scattered cumulus; increased clouds and rain in the afternoon \\
    21b & CON & ACTA & 22 Jul 05--23 & shallow cumulus \\
    22a & GW & IVTA & 23 Jul 02--15 & weak gravity-wave activity \\
    22b & MoBL & IVTA, ACTA & 23 Jul 02--15 & well developed valley winds; deep convection and thunderstorm in the afternoon \\
    22c & CON & ACTA & 23 Jul 02--15 & scattered growing cumulus in the morning; deep convection in the afternoon \\
    23 & CON & ACTA & 24 Jul 05--14 & broken cumulus; convective development \\
    24 & SYN & all & 24 Jul 05--26 Jul 17 & upper-level front associated with cut-off low
flow \\
    \hline
\caption{IOPs during the sEOP, with their start and end time and a preliminary characterization. IOP types are defined in Section~\ref{sec:toc}.}
\label{tab:iops_seop}
\end{longtable}
\end{landscape}

\newpage
\bibliographystyle{apalike} 
\bibliography{toc}

\end{document}